\documentclass[aps,pra,amsmath,amssymb,floatfix,twocolumn,amsmath,superscriptaddress,twocolumn,nofootinbib,tighten,letterpaper]{revtex4-1}
\usepackage[colorlinks,linkcolor=blue,citecolor=blue,urlcolor=blue]{hyperref}
\usepackage{multirow}
\usepackage{bbold}
\usepackage{subfigure}
\usepackage{color}
\usepackage{mathrsfs}
\usepackage{hyperref}
\usepackage[normalem]{ulem}
\usepackage{bm}
\usepackage{pifont}

\usepackage{amsfonts, relsize, color}
\usepackage{graphics}
\usepackage{graphicx}
\usepackage{subfigure}
\usepackage{hyperref}
\usepackage[table]{xcolor}
\usepackage{comment}

\newcommand{\cmark}{\ding{51}}%
\newcommand{\xmark}{\ding{55}}%

\newcommand{\ket}[1]{\left| #1 \right>}

\definecolor{RowColor}{rgb}{0.88,1,0.9}

\begin{document}

\title{Color degeneracy of competing orders near topological defects cores in planar quadratic band touching systems}

\author{Bitan Roy}
\affiliation{Department of Physics, Lehigh University, Bethlehem, Pennsylvania, 18015, USA}

\date{\today}

\begin{abstract}
We study two-dimensional fermionic systems, displaying quadratic band touching in the normal state, in the presence of vortices and skyrmions of insulating and superconducting masses in the ordered phase. A prototypical example of such systems is the Bernal bilayer graphene that supports eight zero-energy modes in the presence of a mass vortex with the requisite U(1) symmetry. Near the vortex core, additional ten masses that close an SO(5) algebra can develop local expectation values by splitting the manifold of zero modes in five and ten different ways by lifting its SO(4) and SU(2) chiral symmetries, respectively. In particular, each SU(2) chiral symmetry can be broken by three distinct copies of chiral-triplet mass orders, giving rise to the notion of the color degeneracy among the competing orders near the vortex core. By contrast, a skyrmion of three anticommuting masses supports additional six masses in its core, and possesses an SU(2) isospin quantum number, besides the usual generalized U(1) charge. Consequently, charge $4e$ Kekul\'e pair density waves can develop in the skyrmion core of N\'eel layer antiferromagnet, while a skyrmion of quantum spin Hall insulator in addition supports an $s$ wave pairing. We also analyze the internal algebra of competing orders in the core of these defects on checkerboard or Kagome lattice that supports only a single copy of quadratic band touching in the normal state.      
\end{abstract}

\maketitle

\vspace{10pt}

\section{Introduction}

The transition between two distinct broken symmetry phases, even though commonly believed to be first-order in nature, can be continuous when two order parameters are related via a chiral rotation (see below). Such unconventional continuous phase transition possibly takes place through proliferation of real space singularities, known as topological defects, when one order resides inside the defect core of the other, giving rise to the notion of competing or dual orders and deconfined criticality~\cite{DQCP-science, DQCP-PRB}. A well studied example of such dual or competing orders involves the N\'{e}el antiferromagnet and valence bond solid in two-dimensional frustrated spin models of insulating systems, respectively breaking the spin rotational and translational symmetries~\cite{DQCP-Sandvik, DQCP-Kaul, DQCP-Nahun-1, DQCP-Nahun-2, lauchli}.

The notion of competing orders becomes more transparent, when they can be described as composite objects of underlying fermionic degrees of freedom. In this respect, massless Dirac fermions, realized in monolayer graphene (MLG), $d$-wave superconductor, and honeycomb Kondo-Heisenberg model, for example, constitute an ideal platform to capture the physics of competing orders~\cite{abanov-weigman, tanaka-hu, senthil-fisher-prb, grover-senthil, hou-mudry-chamon-ryu, seradjeh-moore-franz, ghaemi-ryu-lee, herbut-prl-2010, herbut-isospin, herbut-lu-roy, sachdev-xu-fu, chakravarty, si-goswami, liu-si-goswami}. Namely, in a multicomponent spinor basis (arising from the sublattice or orbital, valley or isospin, and real spin degrees of freedom) ordered phases are represented by Dirac bilinears. Two competing orders are then described by mutually anticommuting Dirac matrices, which when in addition anticommute with the Dirac Hamiltonian, are named \emph{masses}. Naturally, the generators of the chiral rotation between any two competing masses commute with the Dirac Hamiltonian, manifesting its global chiral symmetry~\cite{HJR, Roy-Goswami-Juricic}.

However, representation of ordered phases in terms of Dirac matrices is not limited to the Dirac materials, rather they are quite natural for any multiband systems. Even though mass orders (generators of chiral symmetry) follow universal definition, namely their corresponding matrix operators anticommute (commute) with the normal-state Hamiltonian for effective noninteracting quasiparticles~\cite{HJR, chiralsymm:1, chiralsymm:2, chiralsymm:3, chiralsymm:4}, their explicit form and internal algebra depends on microscopic details. See and compare Table~\ref{tab-2} and Table~\ref{tab-3}, for example. Here we address competing orders that can be found in the core of topological defects, such as vortices and skyrmions, in planar fermionic systems, where the valence and conduction bands in the normal state display biquadratic touching, also known as Luttinger materials. The Bernal stacked bilayer graphene (BLG) is an ideal place to realize such unusual gapless fermionic excitations~\cite{graphene-RMP}. Theoretical studies have shown that quadratic band touching (QBT) in Bernal BLG can be unstable toward the formation of various broken symmetry phases, the exact nature of which depends on a number of microscopic details, such as the relative strength of various finite range components of the Coulomb interactions~\cite{nandkidhore-prl, nandkidhore, fanzhang-polini,vafek-solo, aleiner-2010, vafek, aleiner, fanzhang, scherer-honerkamp, assaad-BLG, cvarma, kharitonov, roy-BLG-QHE, milovanovic, dassarma, wang-2020, szaboroyBLG:2021}, even if that may require a finite interaction couplings~\cite{pujari, janssen, thomaslang}. A number of ordered phases have also been observed in experiments in the presence or absence of external magnetic and displacement electric fields~\cite{freitag-1, lau, freitag-2, freitag-3, nematic-1, nematic-2, veligura-exp, lau-2014}. Therefore, understanding the role of topological defects and competing orders in QBT systems is a timely topic of pressing importance.

Various theoretical works in the recent past have discussed the possibilities of topological defects and dual orders in BLG in the absence~\cite{herbut-lu, moon, sachdev-1} as well as in the presence of magnetic fields~\cite{sondhi, sachdev-2}, and also predicted a charge $4e$ $s$-wave superconductor induced by skyrmions of topological quantum spin Hall insulator (QSHI)~\cite{herbut-lu, moon}. Despite these commendable efforts, the internal algebra of competing orders in the core of topological defects in QBT systems still remains vastly unexplored, and constitutes the central theme of the present work. Here we use the \emph{real} Clifford algebra of anticommuting matrices (all of them are \emph{real}) to address this question~\cite{okubo}.

The most tantalizing outcomes in this context are possibly the following. We find that skyrmions of both N\'eel antiferromagnet and QSHI in BLG, each being constituted by three mutually anti-commuting mass matrices, accommodate charge $4e$ spin-singlet pair-density-waves, assuming two distinct Kekul\'e patterns on honeycomb lattice, by developing their finite local expectation values within the manifold of the bound states localized in the vicinity of the defect core, whereas the skyrmion core of QSHI on the same token additionally sustains an $s$-wave pairing. By contrast, in a Dirac system skyrmion core of antiferromagnet is devoid of any superconducting mass, while that of QSHI only supports $s$-wave pairing~\cite{grover-senthil, herbut-isospin}. On the other hand, all three singlet pairings in BLG support topological QSHI in the mixed phase (supporting vortex-type defects), while the N\'eel order can only be realized in the vortex phase of spin-singlet Kekul\'e superconductors. These local order parameters in the vortex core develop by splitting the subspace of zero energy modes with their eigenvalues $+1$ and $-1$, and filling only the former subset, for example~\cite{ herbut-prl-2010}. Thus, in BLG the competing orders are not unique, giving rise to the notion of a \emph{color degeneracy} (defined precisely below) among them. As lattice-based numerical analyses with nontrivial topological defects of mass orders in Dirac systems have confirmed the formation of induced orders or additional competing masses near the defect core~\cite{hou-mudry-chamon-ryu, ghaemi-ryu-lee, liu-si-goswami}, one expects that the nature of competing orders and their color degeneracy in planar QBT systems can also be tested from similar independent lattice-based numerical analyses. Now we present an extended summary of our main findings.

\subsection{Extended summary of results}~\label{subsec:summary}

The differences in the internal algebra of competing orders in MLG and BLG root into the dispersion of noninteracting fermions, which respectively scales linearly and quadratically with the momentum in these two systems. Consequently, the number of mass matrices that can develop a uniform and isotropic spectral gap at the band touching points via spontaneous lifting of discrete and/or continuous symmetries are 36~\cite{hou-mudry-chamon-ryu} and 28~\cite{roy-classification} in MLG and BLG, respectively, despite both of them possessing the same set of symmetries~\cite{graphene-RMP}.~\footnote{The maximal number of mutually anti-commuting masses is five in MLG~\cite{hou-mudry-chamon-ryu, herbut-isospin}, while that is six in BLG~\cite{roy-classification}.} Also in stark contradistinction to Dirac systems, we show that QBT does not necessarily encounter the fermion doubling, and one can realize a two-component QBT for spinless fermions on two-dimensional, such as checkerboard and Kagome lattices~\cite{sun-fradkin-kivelson}. A simple algebraic proof of this statement is offered in Appendix~\ref{append:quadratic-minimal}. Here, such a realization is named `single-flavored QBT', while the QBTs in Bernal BLG is coined as `valley-degenerate QBT'.

$(1)$ In a single-flavored QBT system, a vortex of any two mutually anticommuting masses [see Table~\ref{tab-2}] hosts two states at precise zero-energy and each of them are two-fold degenerate, yielding a total of four (due to spin) zero-energy states.~\footnote{Such two-fold degeneracy of each zero mode is protected by a pseudo time-reversal symmetry~\cite{herbut-lu}, discussed in Sec.~\ref{subsec:vortexbuldingblock}.} But, three competing mass matrices can split the zero-energy manifold by developing finite expectation values, and they close an SU(2) algebra, see Fig.~\ref{Fig-2}. For example, the zero-energy modes bound to the vortex of an $s$-wave superconductor supports all three components of the QSHI. On the other hand, a skyrmion core of QSHI accommodates the $s$-wave pairing.

$(2)$ A real space vortex of two anticommuting masses with the requisite U(1) symmetry accommodates doubly-degenerate four, thus total eight states at zero energy in a valley-degenerate QBT system. The sub-space of zero-energy states altogether supports ten masses. For example, a vortex of translational symmetry breaking Kekul\'e current order sustains the layer polarized state (1), N\'eel layer antiferromagnet (3), the real (3) and imaginary (3) components of the spin-triplet $f$-wave pairing. Quantities in the parentheses indicate the number of matrices required to describe a particular order, see Table~\ref{tab-3}. Other examples are discussed in Sec.~\ref{sussubsection:valleyQBT_vortex}. Irrespective of these details, the ten masses close an SO(5) algebra.~\footnote{In a Dirac material, such as MLG, the six masses bound to the vortex zero-modes close an SU(2)$\otimes$SU(2)$\cong$SO(4) algebra~\cite{herbut-isospin}.}

\begin{table}[t!]
\begin{tabular}{|c c c c c|}
\hline
{\bf Mass order} & {\bf Matrix} & $I_{uv}$ & $\vec{S}$ & $I_T$ \\
\hline \hline
QAHI & $\tau_0 \otimes \sigma_0 \otimes \alpha_2$ & $-$ & \cmark & $-$ \\
\rowcolor{RowColor}
QSHI & $\tau_3 \otimes \vec{\sigma} \otimes \alpha_2$ & $-$ & \xmark & $+$  \\
$s$-wave pairing & $(\tau_1,\tau_2) \otimes \sigma_0 \otimes \alpha_0$ & $+$ & \cmark & $(+,-)$ \\
\hline
\end{tabular}
\caption{All the mass matrices in a single-flavored quadratic band touching system in a checkerboard lattice~\cite{andras-thesis}, and their transformation under the exchanges of two sublattices ($I_{uv}$), rotation of spin quantization axes ($\vec{S}$), and reversal of time ($I_T$). Here, $+\;(-)$ corresponds to even (odd), and \cmark and \xmark\; reflect whether a mass operator preserves a particular symmetry or not, respectively. Three sets of Pauli matrices $\{ \tau_\mu \}$, $\{ \sigma_\mu\}$, and $\{ \alpha_\mu \}$ operate on the Nambu or particle-hole, spin, and sublattice indices, respectively, with $\mu=0,\cdots, 3$. The real and imaginary components of the $s$-wave pairing appear with $\tau_1$ and $\tau_2$, respectively. Here, QAHI (QSHI) stands for quantum anomalous (spin) Hall insulator. 
}~\label{tab-2} 
\end{table} 

$(3)$ In the vicinity of the vortex core, a set of ten mass terms forming such an SO(5) algebra may acquire local expectation values by splitting the manifold of zero modes at the cost of their SO(4) and SU(2) chiral symmetries. Any set of such ten masses can be organized into five sets of four mutually anticommuting masses, closing an SO(4) algebra [see Fig.~\ref{so4_squares} and Appendix~\ref{append:so5so4}]. Therefore, if the system chooses to split the zero-energy manifold by lifting its SO(4) chiral symmetry, there are five such choices. On the other hand, an SO(5) group has ten SO(3) or SU(2) subgroups that leave ten mutually orthogonal three-dimensional hyperplanes invariant under SO(3) or SU(2) rotations. Thus, zero-energy subspace can also be split by breaking its SU(2) chiral symmetry in ten different ways. But, each set of SU(2) generators rotate between three distinct sets of three mutually anticommuting masses, see Fig.~\ref{SU2_subgroups}. Hence, each SU(2) chiral symmetry of zero modes can be lifted in three different patterns, leading to the notion of the color degeneracy among the competing orders inside the vortex core. For example, either the N\'eel layer antiferromagnet, and the real and imaginary components of triplet $f$-wave pairing can split the manifold of zero modes bound to the vortex of singlet Kekul\'e current orders by spontaneously lifting the SU(2) spin rotational symmetry in all these cases.       

\begin{table*}[t!]
\begin{tabular}{|c c c c c c c c|}
\hline
{\bf Mass order} & {\bf Matrix} & $I_{uv}$ & $I_{K}$ & $\vec{S}$ & $I_{tr}$ & $I_T$ & Symbol \\
\hline \hline
Layer polarized &  $\tau_0 \otimes \sigma_0 \otimes \eta_0 \otimes \alpha_3$ & $-$ & $+$ & \cmark & \cmark & $+$ & LP \\
\rowcolor{RowColor}
Quantum anomalous Hall insulator & $\tau_0 \otimes \sigma_0 \otimes \eta_3 \otimes \alpha_3$ & $-$ & $-$ & \cmark & \cmark & $-$ & QAHI \\
Odd-Kekul\'e charge current & $\tau_0 \otimes \sigma_0 \otimes \eta_2 \otimes \alpha_2$ & $-$ & $-$ & \cmark & \xmark & $-$ & K$_{\rm O}$ \\  
\rowcolor{RowColor}
Even-Kekul\'e charge current & $\tau_3 \otimes \sigma_0 \otimes \eta_1 \otimes \alpha_2$ & $-$ & $+$ & \cmark & \xmark & $-$ & K$_{\rm E}$ \\
\hline
Layer antiferromagnet & $\tau_3 \otimes \vec{\sigma} \otimes \eta_0 \otimes \alpha_3$ & $-$ & $+$ & \xmark & \cmark & $-$ & $\vec{\rm N}$  \\
\rowcolor{RowColor}
Quantum spin Hall insulator & $\tau_3 \otimes \vec{\sigma} \otimes \eta_3 \otimes \alpha_3$ & $-$ & $-$ & \xmark & \cmark & $+$ & $\vec{\rm SH}$ \\
Odd-Kekul\'e spin current & $\tau_3 \otimes \vec{\sigma} \otimes \eta_2 \otimes \alpha_2$ & $-$ & $-$ & \xmark & \xmark & $+$ & $\vec{\rm K}_{\rm O}$ \\  
\rowcolor{RowColor}
Even-Kekul\'e spin current & $\tau_0 \otimes \vec{\sigma} \otimes \eta_1 \otimes \alpha_2$ & $-$ & $+$ & \xmark & \xmark & $+$ & $\vec{\rm K}_{\rm E}$ \\
\hline \hline
$s$-wave pairing & $(\tau_1,\tau_2) \otimes \sigma_0 \otimes \eta_1 \otimes \alpha_1$ & $+$ & $+$ & \cmark & \cmark & $(+,-)$ & $({\rm S}_1,{\rm S}_2)$ \\
\rowcolor{RowColor}
$s$-Kekul\'e pairing & $(\tau_1,\tau_2) \otimes \sigma_0 \otimes \eta_0 \otimes \alpha_0$ & $+$ & $+$ & \cmark & \xmark & $(+,-)$ & $({\rm sK}_1, {\rm sK}_2)$ \\
$p$-Kekul\'e pairing & $(\tau_1,\tau_2) \otimes \sigma_0 \otimes \eta_3 \otimes \alpha_0$ & $+$ & $-$ & \cmark & \xmark & $(+,-)$ & $({\rm pK}_1, {\rm pK}_2)$ \\
\rowcolor{RowColor}
$f$-wave pairing & $(\tau_1,\tau_2) \otimes \vec{\sigma} \otimes \eta_2 \otimes \alpha_1$ & $+$ & $-$ & \xmark & \cmark & $(+,-)$ & $ ( \vec{\rm F}_1,\vec{\rm F}_2 )$ \\
\hline
\end{tabular}
\caption{All the mass matrices in Bernal-stacked bilayer graphene (supporting valley-degenerate quadratic band touching) that anticommute with $\hat{H}^{\rm BLG}_0$, see Eq.~(\ref{hamil-BLG})~\cite{roy-classification}. First eight candidates represent isotropic insulating, and last four to fully and isotropically gapped superconducting states. Among the insulating masses, first four are spin-singlet, while the remaining ones are spin-triplet. From the third to seventh column we display the transformations of these masses under the exchanges of the layers ($I_{uv}$), valleys ($I_{K}$), rotation of the spin quantization axis ($\vec{S}$), U(1) translational symmetry ($I_{tr}$), and reversal of time ($I_T$). The Pauli matrices $\{ \tau_\mu \}$, $\{ \sigma_\mu \}$, $\{ \eta_\mu \}$, and $\{ \alpha_\mu \}$ operate on Nambu or particle-hole, spin, valley, and layer indices, respectively, where $\mu=0,\cdots,3$. Rest of the notations are same as in Table~\ref{tab-2}. Notice that all the masses are rotationally symmetric, leading to a uniform and isotropic gap in the ordered state. 
}~\label{tab-3} 
\end{table*}

$(4)$ In the presence of an underlying skyrmion of three mutually anticommuting masses, there is no bound state at zero energy. But, the sub-space of bound states at finite energies features an SU(2)$\otimes$U(1) chiral symmetry. While the generator of U(1) rotation captures the generalized chiral charge of the skyrmion, the SU(2) generators correspond to its isospin, see Fig.~\ref{Fig-triangle}. Altogether a skyrmion core supports six induced mass orders. The U(1) charge causes rotation among three distinct copies of induced masses, while each SU(2) generator rotates between two distinct flavors of masses. Thus by developing finite expectation value of its charge or isospin quantum number, a skyrmion core can support degenerate flavors of competing induced masses, once again giving rise to the notion of the color degeneracy among competing orders in its core. Consequently, one can construct multiple copies of five mutually anticommuting masses [see Sec.~\ref{subsubsec:skyrmionvalleyQBTalgebra}], the right number to sustain a Wess-Zumino-Witten (WZW) term in $d=2$~\cite{wess-zumino,witten}, after integrating out the fermions~\cite{abanov-weigman, jaroszewicz}. However, due to the color degeneracy of competing orders the skyrmion possesses an induced U(1) charge and SU(2) isospin quantum numbers  (see Fig.~\ref{Fig-triangle}). Consequently, one can construct either a charge-WZW term by developing finite expectation value of the U(1) charge or isospin-WZW term by developing an isospin quantum number (defined more precisely in Sec.~\ref{subsubsec:skyrmionvalleyQBTalgebra}), which can be responsible for continuous and possibly deconfined phase transitions between competing phases that nowadays can also be tested in quantum Monte Carlo simulations~\cite{assaad-QSHI-SC-Natcomm, assaad-mong-1, assaad-mong-2}, for example.

Here we solely focus on the mass orders in QBT systems, which by virtue of producing uniform and isotropic gapped quasiparticle spectra in the ordered phases become energetically most favored at zero and very low temperatures. The microscopic four-fermion interactions that can ultimately stabilize each of these phases in BLG has recently been worked out within the framework of an unbiased renormalization group analysis~\cite{szaboroyBLG:2021}. Also the experimental signatures of each mass order have been discussed in Sec.~IX of Ref.~\cite{roy-classification}. Distinct and unique structure of competing mass orders in the cores of topological defects in QBT systems in comparison to those in Dirac systems (featuring linear band touching in the normal state), which we have highlighted above and Clifford algebraically established later (see Sec.~\ref{sec:cliffordaalgebra}), motivates the present discussion.

\subsection{Organization}

The rest of the paper is organized in the following way. In the next section, we discuss the microscopic models leading to both single-flavored and valley-degenerate QBTs, and all possible mass orders therein. See also Tables~\ref{tab-2} and~\ref{tab-3}. Topological defects, such as vortices and skyrmions, and the bound states in their cores are discussed in Sec.~\ref{sec:topologicaldefects}. Section~\ref{sec:cliffordaalgebra} is devoted to the derivation of the internal algebra among competing orders in the defect cores using the real representation of the Clifford algebra (constituted by real matrices). We support these findings through a plethora of concrete physically relevant examples in Sec.~\ref{sec:examples}, and summarize the results in Sec.~\ref{sec:summary}. Readers only interested in physical examples may wish to skip Secs.~\ref{sec:QBTmassgeneral}-\ref{sec:cliffordaalgebra} containing algebraic structure and internal symmetry of mass orders, and directly go to Sec.~\ref{sec:examples}. Additional discussions are relegated to two appendixes. Specifically, in Appendix~\ref{append:quadratic-minimal} we show that QBT in time-reversal symmetric systems can be devoid of doubler. In Appendix~\ref{append:so5so4}, we review the Lie algebras of SO(5) and SO(4) groups.

\section{Masses in QBT systems}~\label{sec:QBTmassgeneral}

We begin the discussion by considering microscopic models for QBTs.~Unlike the situation in two-dimensional Dirac materials with time reversal symmetry, displaying linear touching of the valence and conduction bands, for which the minimal representation must be four component for spinless fermions (Nielsen-Ninomiya fermion doubling)~\cite{nielsen-ninomiya, herbut-minimalrepresentation}, a two-component QBT can be realized in two-dimensional lattices with finite-range hopping among spinless fermions. Such realizations are compatible with the requirement of the time-reversal symmetry, as shown in Appendix~\ref{append:quadratic-minimal}. Nevertheless, it is also conceivable to realize four-component QBTs in two-dimensional lattices, such as in Bernal stacked BLG in the presence of intralayer nearest-neighbor and interlayer vertical dimer hopping elements. In this system two copies of two-component QBT are realized near two inequivalent corners, also known as the valleys, of the hexagonal Brillouin zone~\cite{graphene-RMP}. Below we write down the low-energy models of these systems and tabulate all possible mass orders therein. See also Table~\ref{tab-2} and~\ref{tab-3}.

\subsection{Single-flavored QBT}~\label{subsec:singleQBT_setup}

The simplest microscopic model, supporting a single copy of QBT can be realized on a checkerboard lattice~\cite{sun-fradkin-kivelson}. To accommodate all possible masses in such a system, we introduce an eight-component Nambu-doubled spinor $\Psi=\left( \Psi_p, \Psi_h \right)^\top$, where $\Psi_p$ and $\Psi_h$ are two four-component spinors, with $\Psi^\top_p=(\Psi_{p,\uparrow}, \Psi_{p,\downarrow})$ and $\Psi^\top_h=(\Psi_{h,\downarrow}, -\Psi_{h,\uparrow})$. The two-component spinors are  
\begin{align}
\Psi^{\top}_{p,\sigma}=\left[ u_{\sigma}, v_{\sigma} \right] ({\mathbf k}) 
\; \text{and} \;
\Psi^{\top}_{h,\sigma}=\left[ u^\dagger_{\sigma}, v^\dagger_{\sigma} \right] (-{\mathbf k}).  
\end{align}
Here $u_\sigma ({\bf k})$ and $v_\sigma ({\bf k})$ correspond to fermion annihilation operators on two sublattices of the checkerboard lattice with momentum ${\mathbf k}$, measured from the band touching $\Gamma=(0,0)$ point, and spin projection $\sigma=\uparrow, \downarrow$. In this basis, the low-energy Hamiltonian near the $\Gamma$ point is 
\begin{eqnarray}~\label{H-nambu-single}
\hat{H}^{\rm SF}_0 =\tau_3 \otimes \sigma_0 \otimes \left[ \alpha_1 d_2 ({\mathbf k}) + \alpha_3 d_1 ({\mathbf k}) \right],
\end{eqnarray}
where 
\begin{equation}\label{formfactor}
d_1 ({\mathbf k})=\frac{k^2_x-k^2_y}{2 m_\ast}, \quad d_2 ({\mathbf k})=\frac{ 2 k_x k_y}{2 m_\ast},
\end{equation} 
and $m_\ast$ has the dimension of mass. Three sets of Pauli matrices $\{ \alpha_\mu \}$, $\{ \sigma_\mu \}$, and $\{ \tau_\mu\}$ operate on the sublattice, spin, and Nambu indices, respectively, where $\mu=0,1,2,3$, and `$\otimes$' represents a direct or tensor product. Throughout, we neglect the particle-hole anisotropy.

The above Hamiltonian ($\hat{H}^{\rm SF}_0$) is invariant under the (1) exchange to two sublattices ($u \leftrightarrow v$), generated by $I_{uv}=\tau_0 \otimes \sigma_0 \otimes \alpha_1$, under which $(k_x,k_y) \to (k_y,k_x)$, (2) reversal of time, generated by the antiunitary operator $I_{T}=\left( \tau_0 \otimes \sigma_2 \otimes \alpha_0 \right) {\mathcal K}$, where ${\mathcal K}$ is the complex conjugation, such that $I^2_{T}=-1$, and (3) rotation of the spin quantization axis, generated by $\vec{S}=\tau_0 \otimes \vec{\sigma} \otimes \alpha_0$.

Various mass orders in this system that uniformly and isotropically gap the QBT point and their transformations under various discrete ($I_{uv}$ and $I_T$) and continuous ($\vec{S}$) symmetries of $\hat{H}^{\rm SF}_0$ are shown in Table~\ref{tab-2}. Altogether, a single-flavored QBT supports three physical masses, namely the quantum anomalous Hall insulator (QAHI), QSHI, and spin-singlet $s$-wave pairing. But, it requires six matrices to describe them~\cite{andras-thesis}. Notice that QAHI only anticommutes with $\hat{H}^{\rm SF}_0$, but commutes with remaining two masses. Hence, for the following discussion on the competing orders, captured in terms of mutually anticommuting masses, the QAHI does not play any role.

\subsection{Valley-degenerate QBTs: Bernal BLG}~\label{subsec:valleyQBT_setup}

Next we focus on the QBTs in Bernal-stacked BLG. Unlike the previous example, BLG accommodates two copies of QBT, yielding the valley degeneracy. The corresponding sixteen-dimensional low-energy Hamiltonian reads 
\begin{equation}\label{hamil-BLG}
\hat{H}^{\rm BLG}_0 =\tau_3 \otimes \sigma_0 \otimes \left[ \left( \eta_0 \otimes \alpha_1 \right) d_1 ({\mathbf k}) + \left( \eta_3 \otimes \alpha_2 \right) d_2 ({\mathbf k}) \right],
\end{equation}
where the newly introduced set of Pauli matrices $\{ \eta_\mu \}$ operate on the valley index. The sixteen-component Nambu-doubled spinor basis is $\Psi=\left( \Psi_p, \Psi_h \right)^\top$, where $\Psi^\top_p=\left( \Psi_{p,\uparrow}, \Psi_{p,\downarrow} \right)$ and $\Psi^\top_h=\left( \Psi_{h,\downarrow},-\Psi_{h,\uparrow} \right)$ are two eight-component spinors. The four-component spinors are 
\begin{align}
\Psi^\top_{p,\sigma} &= \left[u_{+,\sigma}, v_{+,\sigma}, u_{-,\sigma}, v_{-,\sigma} \right]({\mathbf k}), \nonumber \\
\text{and} \; \Psi^\top_{h,\sigma} &= \left[v^\dagger_{+,\sigma}, u^\dagger_{+,\sigma}, v^\dagger_{-,\sigma}, u^\dagger_{-,\sigma} \right] (-{\mathbf k}), 
\end{align} 
where $u_{\pm,\sigma}({\bf k})$ and $v_{\pm,\sigma}({\bf k})$ are the fermionic annihilation operators on two complimentary layers, with Fourier component localized around the nonequivalent valleys at $\pm {\bf K}$, spin projection $\sigma=\uparrow, \downarrow$, and momentum ${\bf k}$, measured from the corresponding valley.

\begin{figure}[t!]
\centering
\includegraphics[width=0.8\linewidth]{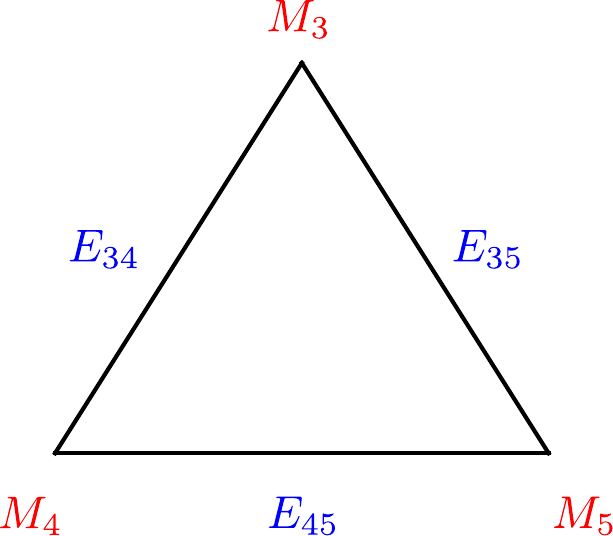}
\caption{Triangle of three mutually anticommuting masses, $M_3$, $M_4$, and $M_5$, that also anticommute with the eight-dimensional vortex Hamiltonian $H^{\rm Nam}_{\rm vor}$ [see Eq.~(\ref{vortex_Nambu})] in a single-flavored quadratic band touching system, occupying its three vertices. Three arms of the triangle represent generator of SU(2) rotations among three masses, generated by $E_{jk}=i M_j M_k$. In particular, each arm causes a U(1) rotation between two mutually anticommuting masses residing at its two ends. 
}~\label{Fig-2}
\end{figure}

The non-interacting Hamiltonian ($\hat{H}^{\rm BLG}_0$) remains invariant under the following symmetries.
\begin{enumerate}
\item Exchange of two layers, generated by $I_{uv}=\tau_0 \otimes \sigma_0 \otimes \eta_0 \otimes \alpha_1$,
\item exchange of two valleys, generated by ${I_K}=\tau_0 \otimes \sigma_0 \otimes \eta_1 \otimes \alpha_0$, 
\item reversal of time, generated by $I_T= \left( \tau_0 \otimes \sigma_2 \otimes \eta_1 \otimes \alpha_0 \right) {\mathcal K}$, such that $I^2_T=-1$,
\item rotation of the spin quantization axis, generated by $\vec{S}=\tau_0 \otimes \vec{\sigma} \otimes \eta_0 \otimes \alpha_0$, and
\item U(1) translational symmetry, generated by $I_{tr}=\tau_3 \otimes \sigma_0 \otimes \eta_3 \otimes \alpha_0$.
\end{enumerate}
It should be notated that the exchange of two layers and valleys are accompanied by momentum reflections $k_y \to - k_y$ and $k_x \to - k_x$, respectively. Therefore, all mass orders can be classified according to their transformation under these symmetries, see Table~\ref{tab-3}.

Altogether Bernal-stacked BLG supports twelve different symmetry breaking mass orders, among which eight (four) are insulators (superconductors). But, one requires 28 matrices to describe them~\cite{roy-classification}. Note that in BLG Kekul\'e valence bond solids~\cite{hou-chamon-mudry-prl} (both spin-singlet and spin-triplet) no longer represent masses. They are replaced by Kekul\'e current orders (symbolically represented by ${\rm K}_{\rm E}, {\rm K}_{\rm O}, \vec{{\rm K}}_{\rm E}$, and $\vec{{\rm K}}_{\rm O}$). Furthermore, two Kekul\'e spin-triplet mass superconductors in the pairing channels in MLG~\cite{roy-herbut-kekule} are replaced by spin-singlet Kekul\'e pairings in BLG (symbolically represented by ${\rm sK}_1, {\rm sK}_2,{\rm pK}_1$, and ${\rm pK}_2$), reducing the number of mass matrices in BLG to 28 from 36 in MLG~\cite{hou-mudry-chamon-ryu}. These differences will play important roles in the internal algebra of competing orders inside the core of topological defects, which we discuss next. In valley-degenerate QBT system, there exists a unique mass order, namely the QAHI, for which the matrix operator commutes with all the other masses. Consequently, QAHI does not play any role in the forthcoming discussion on the competing orders in the topological defect cores of any mass order in this system, which is described in terms of mutually anticommuting masses.

\section{Topological defects}~\label{sec:topologicaldefects}

In this section, we introduce topological defects inside various mass ordered phases. Specifically, we consider vortices and skyrmions, and highlight the structure of the bound states in their cores. This will allow us to construct the internal algebra of competing orders in the core of these defects, discussed in Secs.~\ref{sec:cliffordaalgebra} and~\ref{sec:examples}.

\subsection{Vortex}~\label{subsec:vortexbuldingblock}

The effective single-particle Hamiltonian for a vortex-type point defect involving two mutually anticommuting masses inside the ordered phase in QBT systems assumes the following universal form 
\begin{equation}\label{4d-BLG}
H_{\rm vor}=\gamma_1 \frac{\partial^2_y-\partial^2_x}{2m_\ast} + \gamma_2 \frac{2 \partial_x \partial_y}{2m_\ast} 
           + |m(r)| \left(\gamma_3 C_{n\phi}  + \gamma_5 S_{n\phi} \right),
\end{equation}    
where $C_{n\phi}= \cos (n\phi)$, $S_{n\phi}=\sin (n\phi)$, with $\phi$ as the polar angle and $r$ as the radial coordinate in the $xy$ plane. The radial profile of $m(r)$ is $m(r \to 0) = 0$ and $m(r \to \infty) = m_0$, otherwise arbitrary, where $m_0$ is a constant. For concreteness, we consider a vortex of unit vorticity ($n=1$), as it is the most stable and energetically favored topological defect. Here, $\gamma_j$s are Hermitian matrices satisfying the anticommuting Clifford algebra $\{ \gamma_j, \gamma_k \}=2 \delta_{jk}$, where $\delta_{jk}$ is the Kronecker delta symbol. Therefore, $\gamma_3$ and $\gamma_5$ are the mass matrices. Since $H_{\rm vor}$ involves four mutually anticommuting $\gamma$ matrices, their minimal dimensionality is four. Even though the entire discussion unfolding the internal algebraic structure of competing orders inside the core of vortex defects rests on the above anticommuting relation, for the physically relevant orders, discussed in Secs.~\ref{subsec:summary} and~\ref{sec:examples} we always choose the mass matrices $\gamma_3$ and $\gamma_5$ constituting such defect from the same order, shown in Tables~\ref{tab-2} and~\ref{tab-3}.

It was shown by Herbut and Lu that due to the QBT in the normal phase, the above Hamiltonian describing a unit vortex supports \emph{two} modes at precise zero energy~\cite{herbut-lu}. Such two-fold degeneracy of the zero-energy manifold and rest of the spectrum is assured by an antiunitary operator $J_K=U {\mathcal K}$, where $U$ is a unitary operator, such that $[H_{\rm vor}, J_K]=0$ and $J^2_K=-1$. Therefore, $J_K$ plays the role of a pseudo time-reversal operator. The existence of such antiunitary operator does not depend on the choice of representation of the $\gamma$ matrices. Without any loss of generality, we choose $\gamma_1$ and $\gamma_3$ to be purely imaginary, and $\gamma_1$ and $\gamma_2$ to be purely real. Then, $U=\gamma_1 \gamma_3$. While $J_K$ endows each energy eigenvalue a two-fold degeneracy, existence of the midgap states is guaranteed by the spectral symmetry, generated by an unitary operator $\gamma_0$, such that $\{H_{\rm vor}, \gamma_0 \}=0$. In particular, $\gamma_0=\gamma_1 \gamma_2 \gamma_3 \gamma_5$ is the fifth anticommuting four-dimensional Hermitian $\gamma$ matrix~\cite{gammamatrices-RMP}.

From the above discussion, we can also infer the competing order in the core of the mass vortex. Since $\{ H_{\rm vor},\gamma_0\}=0$, the two zero-energy modes are the eigenstates of $\gamma_0$ with eigenvalue $+1$ or $-1$.~\footnote{If $\{ \ket{\Psi_E} \}$ is a set of eigenstates of a generic Hamiltonian $H$ with eigenvalues $\{E\}$, then $H \ket{\Psi_E}=E \ket{\Psi_E}$. Now if there exists an operator $M$, such that $M^2=I$ and $\{H,M \}=0$, then $H \left( M \ket{\Psi_E} \right)=(-E) \left( M \ket{\Psi_E} \right)$. Therefore, $ M \ket{\Psi_E}$ is also an eigenstate of $H$, but with eigenvalue $-E$, i.e., $ M \ket{\Psi_{\pm E}}=\ket{\Psi_{\mp E}}$. As the eigenvalues of $M$ are $+1$ and $-1$, zero energy states ($E=0$) are eigenstates of $M$ with eigenvalue $+1$ or $-1$.} 
Therefore, filled or empty zero modes yields a finite expectation value of the mass operator $\gamma_0$, i.e. $\langle  \gamma_0 \rangle \neq 0$. Then in the core of the vortex, constituted by the $\gamma_3$ and $\gamma_5$ masses, the system supports their competing mass $\gamma_0$, since $\{ \gamma_0, \gamma_j\}=0$ for $j=1$ and $2$ (thus qualifying as a mass), as well as $j=3$ and $5$ (hence, a competing order). For the minimal model in Eq.~(\ref{4d-BLG}), a finite expectation value of $\gamma_0$ mass places the zero modes at a finite energy. However, for single-flavored and valley-degenerate QBT systems, soon we will find out that competing orders can split the manifold of zero modes symmetrically about the zero energy.

A single-flavored QBT system is described by eight-dimensional Hermitian matrices. Since the effective single-particle Hamiltonian describing a vortex configuration involves only four mutually anticommuting matrices, and their irreducible representation is four-dimensional~\footnote{In a $d$-dimensional irreducible representation all the matrices cannot be cast in block-diagonal form of dimensionality $<d$~\cite{gammamatrices-RMP}.}, the vortex Hamiltonian can always be cast as a orthogonal sum of two copies of $H_{\rm vor}$ from Eq.~(\ref{4d-BLG}). As a result, the core of a mass vortex hosts four zero energy modes. Following the same line of arguments, readers can convince themselves that there exist eight zero energy modes in the core of a mass vortex in valley-degenerate QBT systems. In the following sections, we will discuss the possible competing mass orders and their internal algebra in such higher-dimensional zero-energy manifolds.

\subsection{Skyrmion}

Next we consider a skyrmion of mass orders. It involves three mutually anticommuting mass matrices. The corresponding effective single-particle Hamiltonian is 
\begin{eqnarray}\label{skyr-hamil} 
H_{\rm skyr} &=& \gamma_1 \frac{\partial^2_y-\partial^2_x}{2m_\ast} + \gamma_2 \frac{2 \partial_x \partial_y}{2m_\ast} \nonumber \\
&+& m_1({\bf r}) \gamma_3 + m_2({\bf r}) \gamma_5 + m_3({\bf r}) \gamma_0. 
\end{eqnarray} 
For an underlying skyrmion of unit skyrmion number 
\begin{eqnarray}\label{masstexture}
{\bf m}({\bf r})= m_0 \left( \frac{2r \lambda}{r^2+\lambda^2} C_{\phi}, \frac{2r \lambda}{r^2+\lambda^2} S_\phi, \frac{r^2- \lambda^2}{r^2+\lambda^2} \right),
\end{eqnarray}
where the parameter $\lambda$ determines its core size. Note that $H_{\rm skyr}$ exhausts all five mutually anticommuting four-dimensional $\gamma$ matrices. Therefore, we cannot find any unitary (or antiunitary) matrix that fully anticommutes with $H_{\rm skyr}$ and all states (including the bound ones) reside at finite energies. Nonetheless, these bound states at finite energies continue to enjoy the two-fold degeneracy, as $[H_{\rm skyr}, J_K]=0$, manifesting the pseudo time-reversal symmetry of $H_{\rm skyr}$. Even though the entire discussion unfolding the internal algebraic structure of competing orders inside the core of skyrmion defects rests on the anticommuting relation $\{ \gamma_\mu, \gamma_\nu \}=2 \delta_{\mu \nu}$ for $\mu, \nu=0,1,2,3,5$, for the physically relevant orders, discussed in Secs.~\ref{subsec:summary} and~\ref{sec:examples} we always choose the mass matrices $\gamma_3$, $\gamma_5$, and $\gamma_0$ constituting such defect from the same order, displayed in Tables~\ref{tab-2} and~\ref{tab-3}.

One can render the loss of the spectral symmetry in the following way. Say, we begin with two zero-energy modes (the eigenstates of $\gamma_0$ with eigenvalues $+1$ or $-1$) bound to the core of a vortex, and subsequently introduce the third component of the mass $\gamma_0$ such that it changes sign as we approach the boundary of the system ($ r \to \infty$) from its origin ($r=0$). Therefore, addition of the $\gamma_0$ mass besides constituting a skyrmion texture, pushes the zero-modes bound to a vortex to finite energies. As a direct consequence of the spectral asymmetry, the core of the skyrmion becomes \emph{electrically charged} of charge $+e$ or $-e$ (depending on the sign of $m_0$). The corresponding operator is $Q_{\rm elec}=I_{4}$, where $I_n$ is an $n$-dimensional identity matrix, which is the product of five mutually anticommuting matrices appearing in $H_{\rm skyr}$ 
\begin{equation}
Q_{\rm skyr}=Q_{\rm elec}= \gamma_1 \gamma_2 \gamma_3 \gamma_5 \gamma_0 = I_4, 
\end{equation} 
where $Q_{\rm skyr}$ is the generalized charge of a skyrmion~\cite{herbut-lu-roy}.

For single-flavored and valley-degenerate QBT systems, the Hamiltonian operator in the presence of a background skyrmion of three mutually anticommuting mass orders can be cast as direct or orthogonal sum of two and four copies of $H_{\rm skyr}$, respectively. Such decomposition allows skyrmions to acquire \emph{chiral charges}, while being electrically neutral. Note that any Hermitian matrix operator that commutes with the noninteracting Hamiltonian ($H^{\rm SF}_0$ and $H^{\rm BLG}_0$), such as $\vec{S}$, generates the chiral symmetry of the system, and qualifies as a chiral charge of the skyrmion. On the other hand, any mass operator that anticommutes with $H_{\rm skyr}$ can develop a finite expectation value within the manifold of the bound states localized near the core of a skyrmion.

\section{Real Clifford algebra and competing orders}~\label{sec:cliffordaalgebra}

\begin{figure*}[t!]
\includegraphics[width=1.00\linewidth]{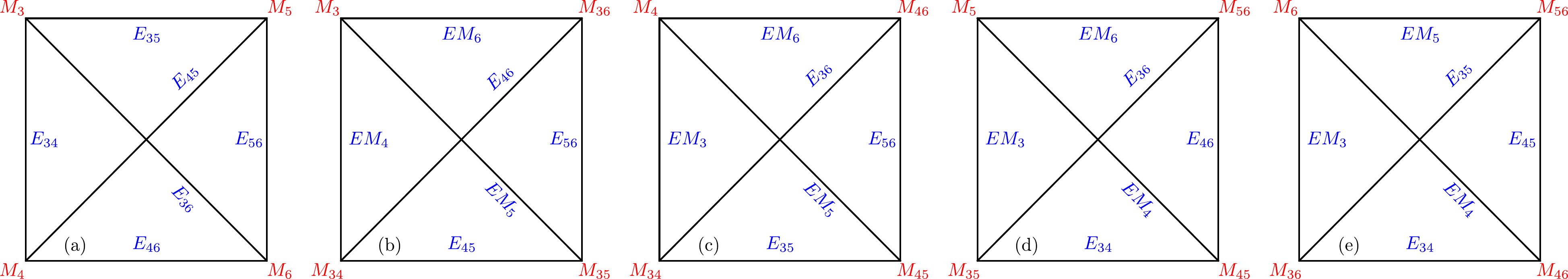}
\caption{Geometric representation of five SO(4) subgroups [see Eq.~(\ref{so4s}) and Appendix~\ref{append:so5so4}], resulting from the generators of SO(5) rotations among ten masses that anticommute with the vortex Hamiltonian $H^{\rm Nam}_{\rm vor}$ in a valley-degenerate QBT system. Four masses (in red) belonging to each SO(4) group reside at the four vertices of a square. Each arm and diagonal of a square stand for the U(1) rotation between two mutually anticommuting masses, residing at its two ends. The generators of all the SO(4) subgroups are depicted in blue.
}~\label{so4_squares}
\end{figure*}

In this section, we derive the internal algebra of competing orders in the core of topological defects using the real representation of Clifford algebra, constituted by all real and mutually anticommuting matrices. In order to describe various insulating and superconducting mass gaps within a unified representation, it is useful to double the number of fermionic components (Nambu doubling), and include both particle and hole in the spinor representation. The resulting massive Nambu Hamiltonian is 
\begin{equation}~\label{eq:ND_massive}
H^{\rm Nam}_m ({\mathbf k}) = H^{\rm Nam}_0 ({\mathbf k}) + m M.
\end{equation}
The kinetic energy part of $H^{\rm Nam}_m ({\mathbf k})$ is given by 
\begin{equation}~\label{eq:ND-kinetic}
H^{\rm Nam}_0 ({\mathbf k})= H_{0} ({\mathbf k}) \oplus H^{\top}_{0} (-{\mathbf k}) \equiv \sum_{j=1,2}\Gamma_j \; d_j ({\mathbf k}), 
\end{equation}
where $\Gamma_j$s, $M$ are eight and sixteen dimensional Hermitian matrices for single-flavored and valley-degenerate QBT systems, respectively. Here, 
\begin{equation}
H_0({\mathbf k}) = \beta_1 \; d_1 ({\mathbf k}) + \beta_2 \; d_2 ({\mathbf k}),
\end{equation}  
and $\beta_j$ are mutually anticommuting four and eight dimensional matrices for these two systems, respectively. The Hermitian matrix $M$ represents a mass order, when it satisfies the anticommutation relation $\{\Gamma_j , M \}=0$ for $j=1,2$. The fully gapped spectra of $H^{\rm Nam}_m ({\bf k})$, namely $\pm \sqrt{[k^2/(2m_\ast)]^2+m^2}$, then extend over positive and negative energies.

By construction the Nambu Hamiltonian $H^{\rm Nam}_m ({\mathbf k})$ preserves the particle-hole symmetry, generated by the antiunitary operator $I_{ph}= \left(\sigma_1 \otimes I_n \right) {\mathcal K}$, and $\{ H^{\rm Nam}_m ({\mathbf k}), I_{ph}\}=0$, with $n=4$ and $8$ for single-flavored and valley-degenerate QBT systems, respectively, and $\sigma_1$ is the real off-diagonal Pauli matrix. Since, $I^{2}_{ph}=+1$, it is always possible to find a representation, known as `Majorana representation', in which $I_{ph}={\mathcal K}$ and $H^{\rm Nam}_m ({\mathbf k})$ is purely imaginary~\cite{herbut-isospin, altland}. In real space representation, the operators $d_j ({\mathbf k} \to -i {\mathbf \nabla})$ are real. Thus two matrices appearing in the kinetic energy ($\Gamma_{1}$ and $\Gamma_2$), as well as any mass matrix ($M$) are imaginary. This is strikingly different from the Dirac system, where due to the linear dependence of $d_j(\mathbf k) \sim k_j$ for $j=1,2$ on spatial components of momentum, $\Gamma_j$'s are purely real. 

\subsection{Single-flavored QBT}

For single-flavored QBT the Nambu Hamiltonian in Eq.~(\ref{eq:ND-kinetic}) is eight dimensional, and $i\Gamma_j$ are purely real. Since $i M$ is also real, we first seek to answer the following question. For eight-dimensional real and mutually anticommuting matrices closing the Clifford algebra $C(p,q)$, what is the maximal value of $q$ in the set of all possible values of $p \geq 0$?~\footnote{The Clifford algebra $C(p,q)$ defines a set of $p+q$ mutually anticommuting real matrices, where $p \; (q)$ of them squares to $+1 \; (-1)$. For example, the maximal number of mutually anticommuting two-dimensional real matrices is three and they are $\sigma_1$, $i \sigma_2$ and $\sigma_3$, where $\sigma_{1,2,3}$ are the standard Pauli matrices. Since $\sigma^2_1=\sigma^2_3=+1$ and $(i \sigma_2)^2=-1$, together they close $C(2,1)$ Clifford algebra of real matrices, with $p=2$ and $q=1$.} The answer is \emph{seven}. They constitute $C(0,7)$ Clifford algebra. Two of them, namely $\Gamma_1$ and $\Gamma_2$, are two imaginary kinetic energy matrices, and $M_j$ are five mutually anticommuting imaginary mass matrices with $j=1,\cdots, 5$~\cite{andras-thesis, okubo}. See also Table~I of Ref.~\cite{herbut-isospin}. There exists another imaginary Hermitian matrix $i \Gamma_1 \Gamma_2$ that anticommutes with the kinetic energy and being purely imaginary it also satisfies the requisite criteria of a mass matrix. But, $i \Gamma_1 \Gamma_2$ commutes with five other mass matrices. Therefore, single-flavored QBT system altogether supports six mass matrices, which we show explicitly in Table~\ref{tab-2}. The $i \Gamma_1 \Gamma_2$ mass can be identified as the QAHI. Next we consider topological defects in such a system.

\subsubsection{Vortex}

If we construct a vortex out of two mutually anticommuting masses, say $M_1$ and $M_2$, according to 
\begin{equation}~\label{vortexmass}
M_{\rm vor} ({\bf x})= |m(r)| \; \left[ M_1 C_\phi + M_2 S_\phi \right],
\end{equation}
then the vortex Hamiltonian, defined as 
\begin{equation}~\label{vortex_Nambu}
H^{\rm Nam}_{\rm vor}=H^{\rm Nam}_0({\mathbf k} \to - i {\boldsymbol \nabla})+ M_{\rm vor} ({\bf x}),
\end{equation}
supports four zero-energy modes, which can be proved in the following way. Note that four mutually anticommuting matrices in $H^{\rm Nam}_{\rm vor}$ close a $C(4,0)$ algebra. Thus the eight-dimensional Hamiltonian $H^{\rm Nam}_{\rm vor}$ can be decomposed in block diagonal form or as orthogonal sum of two identical copies of the four-dimensional Hamiltonian $H_{\rm vor}$, shown in Eq.~(\ref{4d-BLG}). Each such copy hosts two degenerate zero energy states, protected by pseudo time-reversal symmetry~\cite{herbut-lu}. Consequently, the eight dimensional vortex Hamiltonian $H^{\rm Nam}_{\rm vor}$ in Eq.~(\ref{vortex_Nambu}) supports a total of $2 \times 2=4$ states at precise zero energy.

Any matrix that anticommutes with $H^{\rm Nam}_{\rm vor}$ can acquire finite expectation value inside the vortex core by splitting the zero-energy manifold. Such splitting is realized precisely by the mass matrices that anticommute with $H^{\rm Nam}_{\rm vor}$ and their number is only three. They are $M_3,M_4$, and $M_5$, which together close an SU(2) algebra. They can be placed at three vertices of a triangle, see Fig.~\ref{Fig-2}. Three generators of the SU(2) rotations among these three masses are $\left\{E_{34},E_{45},E_{53} \right\}$, where $E_{jk}=i M_j M_k$, which are represented by three arms of the triangle. Also note that $E_{jk}$ commute with $H^{\rm Nam}_{\rm vor}$, thus generating its chiral symmetry.

\subsubsection{Skyrmion}~\label{subsubsec:singleQBT_skyrmion}

Next we proceed to construct a skyrmion out of three mutually anticommuting masses, say $M_1, M_2$, and $M_3$, described by the single-particle Hamiltonian
\begin{eqnarray}~\label{eq:Nambu_skyrmion}
H^{\rm Nam}_{\rm skyr}=H^{\rm Nam}_0({\mathbf k} \to - i {\boldsymbol \nabla}) + \sum^3_{j=1} m_j ({\bf x}) M_j
\end{eqnarray} 
where ${\bf m}({\bf x})$ is given in Eq.~(\ref{masstexture}). Now only one of the three generators of the SU(2) chiral symmetry of $H^{\rm Nam}_{\rm vor}$, namely $E_{45}$, commutes with $H^{\rm Nam}_{\rm skyr}$, and generates its chiral symmetry. This matrix causes U(1) rotations between the remaining two masses $M_4$ and $M_5$ that anticommute with $H^{\rm Nam}_{\rm skyr}$. Therefore, the core of the skyrmion supports these two masses and $E_{45}$ represents its unique generalized charge. It is straightforward to show that  
\begin{equation}
Q_{\rm skyr}=E_{45}=\Gamma_1 \; \Gamma_2 \; M_1 \; M_2 \; M_3,
\end{equation}   
the product of five mutually anticommuting matrices appearing in $H^{\rm Nam}_{\rm skyr}$.~\footnote{This is so because seven mutually anticommuting matrices satisfy the constraint $i \Gamma_1 \Gamma_2 M_1 M_2 M_3 M_4 M_5 \propto I_8$.} In this case, the total of five mutually anti-commuting matrices, yield the right number of matrices in $d=2$ to support a WZW term~\cite{wess-zumino, witten}.

\subsection{Valley-degenerate QBT}~\label{subsec:BLG_Clifford}

As $\hat{H}^{\rm BLG}_0$ in Eq.~\eqref{hamil-BLG} is a sixteen-dimensional Hermitian operator, we initiate the discussion on competing phases in valley-degenerate QBT systems by asking the following question. For sixteen-dimensional real and mutually anticommuting matrices closing a Clifford algebra $C(p,q)$, what is the maximal value of $q$ in the set of all possible $p \geq 0$? The answer is \emph{eight}. They constitute $C(0,8)$ Clifford algebra~\cite{roy-classification}. Two of them can be used to define the noninteracting Hamiltonian in terms of the imaginary matrices $\Gamma_1$ and $\Gamma_2$. The remaining six matrices are the mutually anticommuting mass matrices $M_j$ with $j=1,\cdots,6$. Altogether one can construct 28 imaginary mass matrices in valley-degenerate QBT systems, see Table~\ref{tab-3}. Among these 28 mass matrices, there exists one imaginary mass, namely $i \Gamma_1 \Gamma_2$, which anticommutes with the noninteracting Hamiltonian, but commutes with rest of the 27 masses. It is identified as the QAHI, and does not play any role in the forthcoming discussion on competing orders near the core of topological defects, captured in terms of mutually anticommuting masses.

\begin{figure*}[t!]
\includegraphics[width=1.00\linewidth]{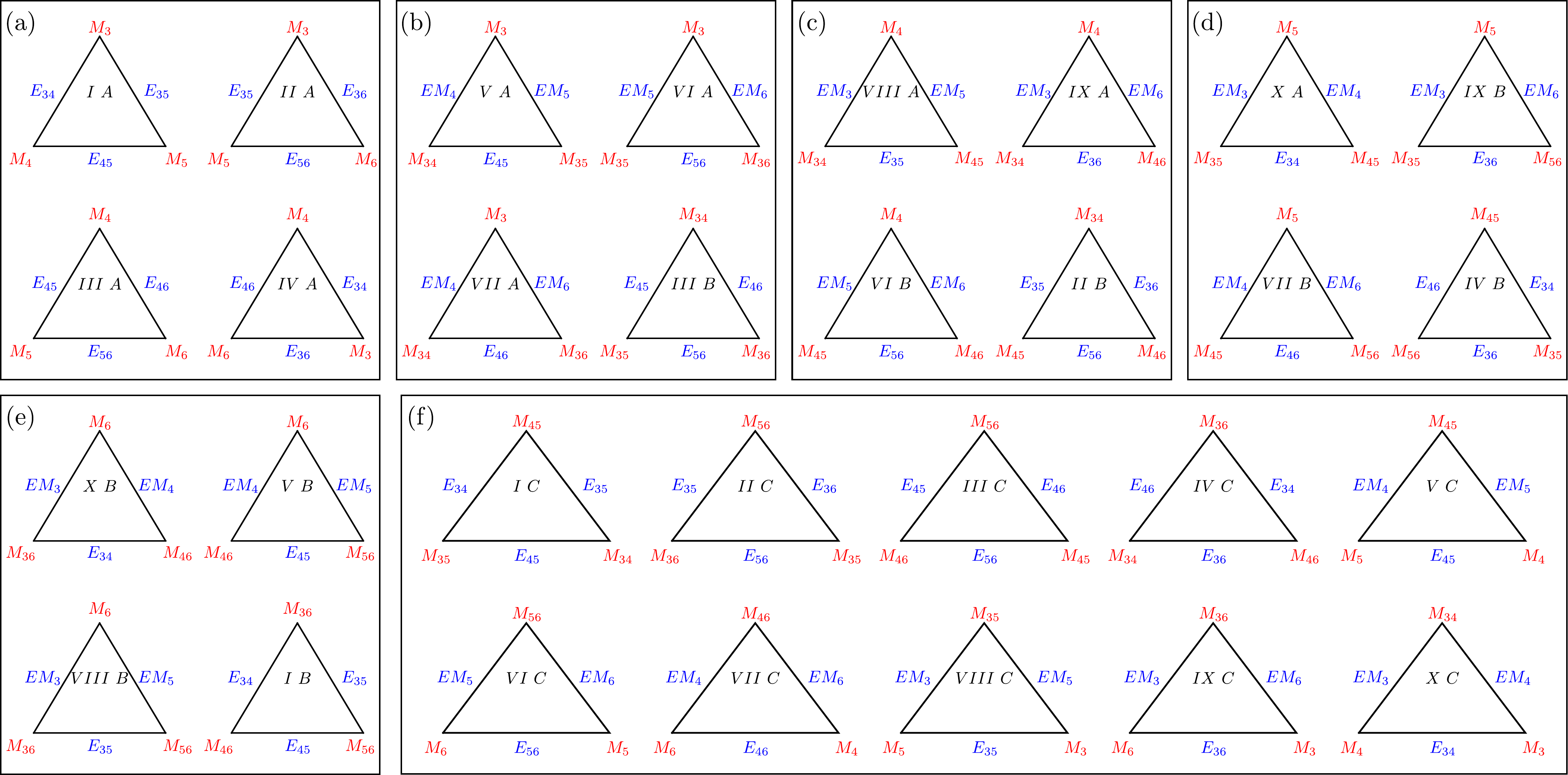}
\caption{(a)-(e): Four SU(2) subgroups (each represented by a triangle, see Fig.~\ref{Fig-2}), resulting from the corresponding SO(4) subgroup [see Fig.~\ref{so4_squares}] of the original SO(5) group of competing masses at the vortex core of a valley-degenerate quadratic band touching system. Each arm of a triangle represents a U(1) rotation between two mutually anticommuting masses (red), residing at its two ends. The corresponding generator is shown in blue. Each set of SU(2) generators $j(=I, \cdots, X)$ [see Eq.~(\ref{eq:su2generator})] rotate between three distinct flavors of three mutually anticommuting masses, occupying the vertices of these triangles $j\alpha$, with $\alpha=A, B$, and $C$, yielding the color degeneracy among the competing chiral triplet masses near the vortex core. The SU(2) triangles $jC$ are shown in $(f)$. 
}~\label{SU2_subgroups}
\end{figure*}

\subsubsection{Vortex}~\label{sussubsection:valleyQBT_vortex}

First we focus on a vortex constituted by two mutually anticommuting mass matrices $M_1$ and $M_2$, following the protocol in Eqs.~(\ref{vortexmass}) and (\ref{vortex_Nambu}). But, now all matrices ($\Gamma_1, \Gamma_2, M_1, M_2$), appearing in these two equations, are sixteen-dimensional. Since these four matrices satisfy $C(4,0)$ algebra, $H^{\rm Nam}_{\rm vor}$ can be cast as an orthogonal sum of four copies of $H_{\rm vor}$, see Eq.~(\ref{4d-BLG}). Consequently, $H^{\rm Nam}_{\rm vor}$ supports eight zero-energy modes.

Any operator, say $X$, that anti-commutes with $H^{\rm Nam}_{\rm vor}$ can acquire a finite expectation value by splitting the subspace of the zero-energy states. To establish the internal structure of such competing orders we need to search for all imaginary matrices $X$ that satisfy the anticommutation relations $\{ X, \Gamma_j \}=\{X,M_j\}=0$, for $j=1,2$. One can immediately find at least four candidates for $X$, namely $M_3$, $M_4$, $M_5$, and $M_6$. However, they do not exhaust all possibilities for $X$. In terms of four imaginary matrices appearing in $H^{\rm Nam}_{\rm vor}$, we can define another Hermitian matrix
\begin{equation}\label{Ematrix}
E= \Gamma_1 \Gamma_2 M_1 M_2.
\end{equation}	
Even though $\{H^{\rm Nam}_{\rm vor},E \}=0$, by construction $E$ is a real matrix. So, $E$ is not a mass matrix. Nevertheless, we can define the following six imaginary Hermitian matrices
\begin{equation}~\label{newmass}
M_{jk} = i E M_j M_k,
\end{equation} 	
where $3 \leq j,k \leq 6$, but with $j \neq k$ and $j>k$, which anticommute with $H^{\rm Nam}_{\rm vor}$. Hence, altogether there are ten mass matrices that anticommute with $H^{\rm Nam}_{\rm vor}$. Any one of them can acquire finite expectation value by splitting the eight-dimensional subspace of zero energy states.

In order to demonstrate the two-fold degeneracy of the zero-energy manifold, we search for all possible candidates for the sixteen-dimensional unitary operator $U$, such that we can define the pseudo time-reversal operator $J_K=U {\mathcal K}$, satisfying $J^2_K=-1$ and $[H_V, J_K]=0$. Since $\Gamma_1, \Gamma_2, M_1, M_2$ are imaginary and $J^2_K=-1$, the imaginary unitary operator $U$ must satisfy $\left\{ H_V, U \right\}=0$. Due to the enlarged dimensionality of $H^{\rm Nam}_{\rm vor}$, in fact there are ten possible choices of $U$, given by the set 
\begin{eqnarray}
U \in \big\{ M_3, M_4, M_5, M_6, M_{34}, M_{35}, M_{36}, M_{45}, M_{46}, M_{56} \big\}. \nonumber 
\end{eqnarray}   
Therefore, any one of the ten masses that anticommutes with $H^{\rm Nam}_{\rm vor}$ can be a candidate for $U$. Since all mass matrices are Hermitian and imaginary, $J^2_K=-1$ by construction. If one of them, say $M_3$, acquires local expectation value ($m_3$) near the vortex core, there are still six candidates for $U$, namely $M_j$ and $M_{3j}$ with $j=4,5,6$, such that $\left[ J_K, H^{\rm Nam}_{\rm vor}+ m_3 M_3 \right]=0$. Therefore, split (by one of the mass orders) manifold of zero energy modes continues to enjoy the two-fold degeneracy.

A question arises quite naturally. What is the internal algebra among these 10 competing masses? Notice each member of the set of 10 masses matrices, say $M_3$, anticommutes with 6 other masses (namely, $M_4,M_5,M_6, M_{34}, M_{35}$, and $M_{36}$), and commutes with 3 other masses (namely, $M_{45},M_{46}$, and $M_{56}$). Such an algebra is the defining property of an SO(5) group, constituted by product matrices. Therefore, 10 masses that can develop finite expectation value within the zero-energy subspace close an SO(5) algebra. By contrast, in a Dirac system (such as MLG) six competing mass orders in the vortex core satisfy SU(2)$\otimes$SU(2) algebra~\cite{herbut-isospin}, whose Lie algebra is isomorphic to that of SO(4). The 10 generators of SO(5) rotations (each of them causing U(1) rotation between two specific mutually anticommuting masses) are given by 
\allowdisplaybreaks[4]
\begin{align}~\label{eq:so5generator}
{\mathcal G} \in \big\{ EM_3, EM_4, EM_5, EM_6, \nonumber \\
E_{34}, E_{35}, E_{36}, E_{45}, E_{46}, E_{56}  
\big\},
\end{align}  
where $E_{jk}=i M_j M_k$. Each generator anticommutes (commutes) with 6 (3) other generators, and they close an SO(5) algebra. See also Appendix~\ref{append:so5so4} for details.

An SO(5) group has five SO(4) subgroups that leave five mutually orthogonal four-dimensional hyperplanes invariant under SO(4) rotations. Between any two of them there exists three common generators, precisely the number of three-dimensional common hyperplanes between two four-dimensional subspaces of a five-dimensional sphere. The generators of each SO(4) subgroups are shown in blue in Fig.~\ref{so4_squares}, and in Appendix~\ref{append:so5so4} we explicitly show that each each of them satisfies SO(4)$\cong$SU(2)$\otimes$SU(2) algebra. In addition, one can construct the following five `four-tuplets' of four mutually anticommuting masses belonging to the SO(4) subgroups
\allowdisplaybreaks[4]
\begin{eqnarray}~\label{so4s}
(a) &\equiv&  \left\{ M_3, M_4, M_5, M_6 \right\}, \nonumber \\
(b) &\equiv&  \left\{ M_3, M_{34}, M_{35}, M_{36} \right\}, \nonumber \\
(c) &\equiv&  \left\{ M_4, M_{34}, M_{45}, M_{46} \right\}, \nonumber \\
(d) &\equiv&  \left\{ M_5, M_{35}, M_{45}, M_{56} \right\}, \nonumber \\
\text{and} \; (e) &\equiv& \left\{ M_6, M_{36}, M_{46}, M_{56} \right\}. 
\end{eqnarray}
In Fig.~\ref{so4_squares} masses are shown in red, and four masses from each SO(4) subgroup reside at the vertices of a square. Four masses belonging to any SO(4) subgroup are mutually anticommuting which reconciles with the fact that the maximal number of mutually anticommuting mass matrices is six in BLG. Therefore, if the system chooses to split the zero-energy manifold by breaking SO(4) chiral symmetry of $H^{\rm Nam}_{\rm vor}$, it can be accomplished in five different patterns.

On the other hand, an SO(5) group has ten SO(3) or SU(2) subgroups that leave ten mutually orthogonal three-dimensional hyperplanes invariant under SO(3) or SU(2) rotations.~\footnote{Ten such specific SU(2) subgroups can be found in the following way. Each SO(4) subgroup yields two SU(2) subgroup, see Appendix~\ref{append:so5so4}. Hence, five SO(4) subgroups give ten SU(2) subgroups. Since, the Lie algebras of the SU(2) and SO(3) groups are isomorphic, each SO(4) subgroup has two SO(3) subgroups.} Their generators are the following
\allowdisplaybreaks[4]
\begin{alignat}{2}~\label{eq:su2generator}
I   &\equiv \left( E_{34}, E_{35}, E_{45} \right), \quad && II   \equiv \left( E_{35}, E_{36}, E_{45} \right), \nonumber \\ 
III &\equiv \left( E_{45}, E_{46}, E_{56} \right), \quad && IV   \equiv \left( E_{46}, E_{34}, E_{36} \right), \nonumber \\
V   &\equiv \left( E M_4,  E M_5,  E_{45} \right), \quad && VI   \equiv \left( E M_5,  E M_6,  E_{56} \right), \nonumber \\
VII &\equiv \left( EM_4,   E M_6,  E_{46} \right), \quad && VIII \equiv \left( E M_3,  E M_5,  E_{35} \right), \nonumber \\
IX  &\equiv \left( E M_3,  E M_6,  E_{36} \right), \quad && X    \equiv \left( E M_3,  E M_4,  E_{34} \right). 
\end{alignat} 
For any $j=I, \cdots, X$, three SU(2) generators ($A_\alpha$s) satisfy the group algebra $[A_\alpha, A_\beta]=i \epsilon_{\alpha \beta \delta} A_\delta$, where $\epsilon_{\alpha \beta \delta}$ is the fully antisymmetric Levi-Civita symbol. As shown in Fig.~\ref{SU2_subgroups}, each set of SU(2) generators can rotate between three distinct flavors of three mutually anticommuting masses, occupying the vertices of three triangles $j\alpha$, where $\alpha=A,B$, and $C$. Therefore, if the system chooses to split the zero-energy manifold by breaking its SU(2) chiral symmetry, there are ten choices ($j=I, \cdots, X$). And each SU(2) chiral symmetry can be broken by three flavors ($jA, jB$ and $jC$) of chiral-triplet masses. Such extra three-fold degeneracy among triplet mass orders is termed here as the \emph{color degeneracy} of competing orders. In Sec.~\ref{subsubsection:vortex_valleyQBT_examples} we show some explicit examples of such color degeneracy.

\subsubsection{Skyrmion}~\label{subsubsec:skyrmionvalleyQBTalgebra}

\begin{figure}[t!]
\centering
\includegraphics[width=1.00\linewidth]{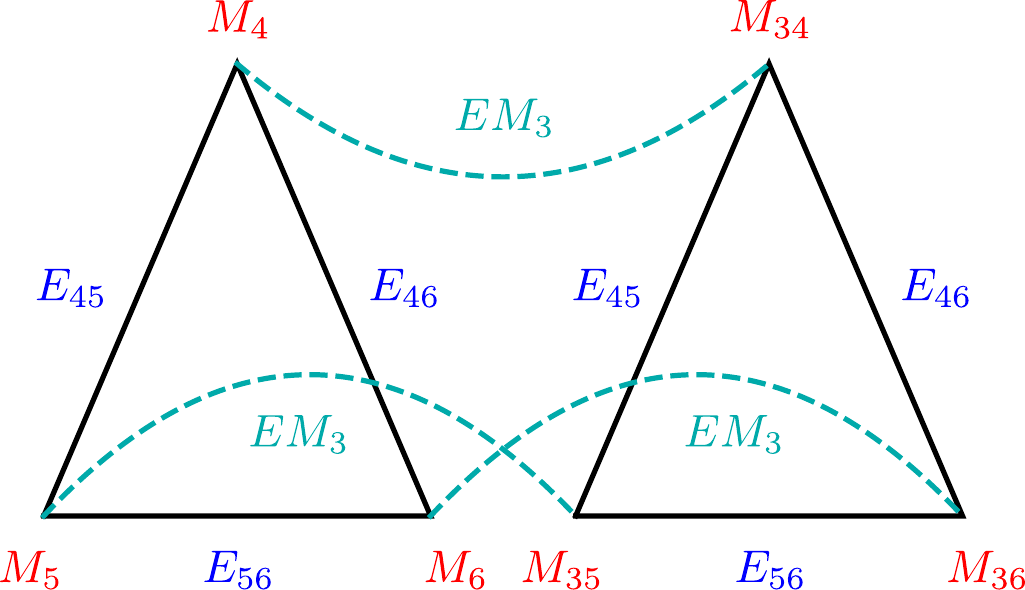}
\caption{Two sets of three mutually anticommuting masses (red) that also anticommute with $H^{\rm Nam}_{\rm skyr}$ and develop finite expectation values in the skyrmion core of a valley-degenerate quadratic band touching system, are placed at three vertices of two triangles. The SU(2) rotations in both triangles are generated by $\left\{E_{45}, E_{56}, E_{64} \right\}$ (represented by the arms of the triangles and are shown in blue), yielding the isospin quantum number of the skyrmion. The dotted lines represent the inter-triangle U(1) rotations, generated by $EM_3$ (the charge of skyrmion), among three pairs of two identical vertices belonging to two triangles, yielding the generalized charge of the skyrmion. See Fig.~\ref{Fig-2} for the details on a geometric representation of an SU(2) group by a triangle.  
}~\label{Fig-triangle}
\end{figure}

Now we focus on a skyrmion with $\Gamma_1$, $\Gamma_2$, $M_1$, $M_2$, and $M_3$ as sixteen-dimensional imaginary Hermitian matrices entering the corresponding effective single-particle Hamiltonian [see Eq.~(\ref{eq:Nambu_skyrmion})]. In the presence of an underlying skyrmion there is no bound state at zero energy. But, all the bound states at finite energies still possess two-fold degeneracy, guaranteed by the pseudo time-reversal operator $J_K= U {\mathcal K}$, with $U \in \left\{ M_4, M_5, M_6, M_{34}, M_{35}, M_{36} \right\}$ that commutes with $H^{\rm Nam}_{\rm skyr}$.

There are two sets of three mutually anticommuting masses that close SU(2) algebra and also anticommute with $H^{\rm Nam}_{\rm skyr}$. They are given by 
\begin{eqnarray}
\left\{ M_4,\; M_5,\; M_6 \right\} \:\: \mbox{and} \: \: \left\{ M_{34},\; M_{35},\; M_{36} \right\}, \nonumber  
\end{eqnarray}
which can be placed at three vertices of two different triangles, see Fig.~\ref{Fig-triangle}. The generators of SU(2) rotations $\left( E_{45}, E_{56}, E_{64} \right)$ are, however, identical for two SU(2) triangles. In addition to the intra-triangle SU(2) symmetry, there exist an inter-triangle U(1) symmetry, generated by 
\begin{equation}
Y=E M_3=\Gamma_1 \Gamma_2 M_1 M_2 M_3,
\end{equation} 
that rotates between two masses residing at identical vertices of these two triangles. The generator of the U(1) symmetry $Y$ is the product of five mutually anticommuting matrices appearing in $H^{\rm Nam}_{\rm skyr}$. As $[H^{\rm Nam}_{\rm skyr}, E_{ij}]=[H^{\rm Nam}_{\rm skyr},Y]=0$, and $[E_{ij},Y]=0$, the bound states in the core of the skyrmion possesses SU(2)$\otimes$U(1) chiral symmetry, which gets broken by the induced competing masses. While $Y$ determines the generalized charge of the skyrmion ($Q_{\rm skyr}$), three generators of the SU(2) rotations correspond to its \emph{isospin}. The notion of the generalized charge is also germane in Dirac~\cite{herbut-lu-roy} and single-flavored QBT systems [see Sec.~\ref{subsubsec:singleQBT_skyrmion}]. But, `isospin' quantum number of the skyrmion is unique to valley-degenerate QBT systems such as the Bernal-stacked BLG.

Notice that the U(1) charge of the skyrmion rotates between three pairs of distinct induced masses, residing at any two equivalent vertices of two triangles, while each generator of isospin SU(2) symmetry rotates between two copies of distinct induced masses, residing at the end of identical arms of two triangles, see Fig.~\ref{Fig-triangle}. Therefore, the induced U(1) or SU(2) quantum number of skyrmions due to the formation of competing mass orders at its core gives rise to the color degeneracy among the competing orders in its core, which we exemplify in Sec.~\ref{subsubsection:skyrmion_valleyQBT_examples}.

In terms of the masses related by the U(1) rotation, generated by $Q_{\rm skyr}=Y$, we can define three copies of five-tuplet of mutually anticommuting masses 
\allowdisplaybreaks[4]
\begin{eqnarray}
{\mathcal T}^{\rm charge}_1 &=& \left\{ M_1, M_2, M_3, M_4, M_{34} \right\}, \nonumber \\
{\mathcal T}^{\rm charge}_2 &=& \left\{ M_1, M_2, M_3, M_5, M_{35} \right\}, \nonumber \\
\text{and} \:\: {\mathcal T}^{\rm charge}_3 &=& \left\{ M_1, M_2, M_3, M_6, M_{36} \right\}. 
\end{eqnarray}
The existence of five mutually anticommuting masses gives the right number to support WZW term in $d=2$~\cite{wess-zumino, witten}. Here such a topological term is named charge-WZW, as three sets of mass orders (last two entries in each of ${\mathcal T}^{\rm charge}_j$ for $j=1,2,3$) develop within the skyrmion core by inducing finite expectation value of its U(1) charge $Q_{\rm skyr}$. Due to the color degeneracy, the same charge-WZW term can arise for three copies of induced masses. One can also construct a WZW term from the following two five tuplets of anticommuting masses  
\allowdisplaybreaks[4] 
\begin{eqnarray}
{\mathcal T}^{\rm isospin,1}_1 &=& \left\{ M_1, M_2, M_3, M_4, M_5 \right\} \nonumber \\
\text{and} \:\: {\mathcal T}^{\rm isospin,1}_2 &=& \left\{ M_1, M_2, M_3, M_{34}, M_{35} \right\},
\end{eqnarray}
where the U(1) rotation between the induced masses, namely $(M_4, M_5)$ (in ${\mathcal T}^{\rm isospin,1}_1$) and $(M_{34},M_{35})$ (in ${\mathcal T}^{\rm isospin,1}_2$), is generated by one of the generators of isospin SU(2) symmetry, namely $E_{45}$. Accordingly, such WZW term is coined isospin-WZW. Therefore, the same isospin-WZW term can arise for two copies of five tuplets of masses, once again manifesting the color degeneracy of competing orders in the skyrmion core. Isospin-WZW terms can also be derived for the induced masses, related via U(1) rotation generated by $E_{46}$ and $E_{56}$, the remaining two generators of SU(2) isospin symmetry of skyrmion. The WZW term in believed to be responsible for continuous (and possibly deconfined) quantum phase transition. Therefore, in valley-flavored QBT systems such an unconventional quantum phase transition can take the system to a variety of competing broken symmetry phases, due to their color degeneracy in the skyrmion core. As the previous attempts of constructing the WZW term did not take into account the color degeneracy of the competing order~\cite{sachdev-1, sachdev-2}, a detailed analysis of it is left for a future investigation.

\section{Examples}~\label{sec:examples}

Upon establishing the internal algebra of competing orders in the core of topological defects (vortices and skyrmions) of mass orders, we now discuss some physically pertinent examples for both single-flavored and valley-degenerate QBT systems that complement the discussion in the introductory section~\ref{subsec:summary}.

\subsection{single-flavored QBT}

To this end we refer to Table~\ref{tab-2} for all masses in this system and Sec.~\ref{subsec:singleQBT_setup} for the corresponding definition of the eight-component spinor. First, we consider a vortex of easy-plane components of QSHI. Therefore, in Eq.~(\ref{vortexmass})
\begin{equation}
(M_1, M_2)=\tau_3 \otimes (\sigma_1, \sigma_2) \otimes \alpha_2, \nonumber 
\end{equation} 
and three mutually anticommuting masses that also anticommutes with $H^{\rm Nam}_{\rm vor}$ are the easy-axis QSHI, and real and imaginary components of the singlet $s$-wave pairing. Any one of these three masses can split the zero-energy manifold of $H^{\rm Nam}_{\rm vor}$ [see Eq.~(\ref{vortex_Nambu})], and acquire a finite expectation value. On the other hand, a vortex inside the $s$-wave paired state, i.e. when 
\begin{equation}
(M_1, M_2)= (\tau_1, \tau_2) \otimes \sigma_0 \alpha_0, \nonumber 
\end{equation}
supports all three components of the QSHI inside its core.

Next we consider a skyrmion of QSHI. The generalized charge of such a skyrmion 
\begin{eqnarray}
Q_{\rm skyr} =\tau_3 \otimes \sigma_0 \otimes \alpha_0,
\end{eqnarray}
is the standard electric charge $Q_{\rm elec}$ in the Nambu doubled basis. Note that $Q_{\rm skyr}$ generates a U(1) rotation between the real and imaginary components of $s$-wave pairing. Therefore, core of the skyrmion of QSHI supports $s$-wave pairing, and a vortex of $s$-wave superconductor allows the local formation of QSHI in single-flavored QBT systems.

\subsection{Valley-degenerate QBT}

Next we discuss the competing phases in the core of vortices and skyrmions in Bernal-stacked BLG, a valley-degenerate QBT system. Readers should consult Table~\ref{tab-3} for sixteen-dimensional representation of the all masses, and Sec.~\ref{subsec:valleyQBT_setup} for the definition of the spinor basis.

	 \subsubsection{Vortex}~\label{subsubsection:vortex_valleyQBT_examples}

		Due to a large number of masses, one can construct a myriad of vortices out of any two mutually anticommuting masses shown in Table~\ref{tab-3}. However, we restrict the discussion to some physically pertinent situations.

		(1) Vortex of spin-singlet Kekul\'e currents ($M_1={\rm K}_{\rm E}$ and $M_2={\rm K}_{\rm O}$): Ten masses that anticommute with ${\rm K}_{\rm E}$ and ${\rm K}_{\rm O}$ are the layer polarized (LP) and layer antiferromagnet ($\vec{{\rm N}}$) states, and spin-triplet $f$-wave superconductor ($\vec{{\rm F}}_1, \vec{{\rm F}}_2$). In this case, the corresponding four-tuplets of mutually anticommuting masses transforming a vector under the SO(4) rotations are  
			\allowdisplaybreaks[4]
	 \begin{eqnarray}
&&\left\{ {\rm LP}, {\rm F}^{1}_1, {\rm F}^{2}_1, {\rm F}^{3}_1 \right\}, \; \left\{ {\rm LP}, {\rm F}^{1}_2, {\rm F}^{2}_2, {\rm F}^{3}_2 \right\}, \; \left\{ {\rm N}^1, {\rm N}^2, {\rm F}^{3}_1, {\rm F}^{3}_2 \right\}, \nonumber\\
&&\left\{ {\rm N}^2, {\rm N}^3, {\rm F}^{1}_1, {\rm F}^{1}_1 \right\}, \; \text{and} \;
\left\{ {\rm N}^3, {\rm N}^1, {\rm F}^{2}_1, {\rm F}^{2}_2 \right\}. \nonumber
\end{eqnarray}
If the zero-energy manifold gets split by spontaneously breaking the SU(2) spin rotational symmetry ($\vec{S}$), it can be accomplished by nucleating either the N\'eel layer anti-ferromagnet ($\vec{{\rm N}}$), real ($\vec{{\rm F}}_1$) or imaginary ($\vec{{\rm F}}_2$) components of the spin-triplet $f$-wave pairing, representing the color degeneracy of competing orders in the vortex core.

   (2) Vortex of easy-plane layer anti-ferromagnet ($M_1={\rm N}^1$ and $M_2={\rm N}^2$) supports singlet Kekul\'e currents (${\rm K}_{\rm E}$ and ${\rm K}_{\rm O}$), Kekul\'e pair density waves (${\rm sK}_1, {\rm sK}_2, {\rm pK}_1$, and ${\rm pK}_2$), and the easy-axis component of QSHI (${\rm SH}^3$), layer anti-ferromagnet (${\rm N}^3$), and $f$-wave pairing (${\rm F}^3_1$ and ${\rm F}^3_2$). Five SO(4) subgroups of competing masses are 
	\allowdisplaybreaks[4]
\begin{eqnarray}
&&\left\{ {\rm N}^3, {\rm K}_{\rm E}, {\rm sK}_1, {\rm sK}_2 \right\}, \left\{ {\rm N}^3, {\rm K}_{\rm O}, {\rm pK}_1, {\rm pK}_2 \right\}, \big\{ {\rm F}^{3}_{1}, {\rm F}^{3}_{2}, {\rm K}_{\rm O}, \nonumber \\ 
&& {\rm K}_{\rm E} \big\}, \left\{ {\rm SH}^{3}, {\rm F}^{3}_{2}, {\rm pK}_2, {\rm sK}_1 \right\}, \; \text{and} \;
\left\{ {\rm SH}^{3}, {\rm F}^{3}_{2}, {\rm pK}_1, {\rm sK}_2 \right\}. \nonumber
\end{eqnarray}
In contrast to a Dirac system (such as MLG), the vortex core of easy-plane N\'eel layer antiferromagnet supports spin-singlet pair-density-waves in Bernal BLG.

   (3) A vortex in the easy-plane of QSHI ($M_1={\rm SH}^1$ and $M_2={\rm SH}^2$) supports easy-axis layer anti-ferromagnet (${\rm N}^3$), QSHI (${\rm SH}^3$), and Kekul\'e spin-currents (${\rm K}^3_{\rm E}$ and ${\rm K}^3_{\rm O}$), $s$-wave pairing (${\rm S}_1$ and ${\rm S}_2$), and Kekul\'e pair density waves (${\rm sK}_1, {\rm sK}_2, {\rm pK}_1$, and ${\rm pK}_2$). The associated four-tuplets of mutually anticommuting masses are 
	\allowdisplaybreaks[4]
\begin{eqnarray}
&& \left\{ {\rm SH}^{3}, {\rm S}_1, {\rm pK}_1, {\rm sK}_2 \right\}, \left\{ {\rm SH}^{3}, {\rm S}_2, {\rm pK}_2, {\rm sK_1} \right\}, \big\{{\rm N}^3, {\rm K}^{3}_{\rm E}, \nonumber \\
&& {\rm pK}_1, {\rm pK}_2 \big\}, \big\{ {\rm N}_3, {\rm K}^{3}_{\rm O}, {\rm sK}_1, {\rm sK}_2 \big\}, \; \text{and} \;
\left\{ {\rm K}^{3}_{\rm E}, {\rm K}^{3}_{\rm O}, {\rm S}_1, {\rm S}_2 \right\}. \nonumber
\end{eqnarray}
In contrast to a similar situation in MLG, where the vortex zero modes only support the $s$-wave pairing, in Bernal BLG they can additionally accommodate translational symmetry breaking Kekul\'e pairings.

So far, we discussed vortices in various insulating phases of BLG, discerning sufficient differences with their counterparts in MLG. Next we discuss vortices of superconducting masses in this system.

(4) First we consider a vortex of $s$-wave pairing, with $M_1={\rm S}_1$ and $M_2={\rm S}_2$. It supports layer polarized state (LP), QSHI ($\vec{{\rm SH}}$), and Kekul\'e spin-currents ($\vec{{\rm K}}_{\rm E}$ and $\vec{{\rm K}}_{\rm O}$). The four-tuplets of mutually anticommuting masses in this case are 
\allowdisplaybreaks[4]
\begin{eqnarray}
&& \left\{ {\rm SH}^1, {\rm SH}^2, {\rm K}^{3}_{\rm O}, {\rm K}^{3}_{\rm E} \right\}, \left\{ {\rm SH}^2, {\rm SH}^3, {\rm K}^{1}_{\rm O}, {\rm K}^{1}_{\rm E} \right\}, \big\{ {\rm SH}^3, {\rm SH}^1, \nonumber \\ 
&&  {\rm K}^{2}_{\rm O}, {\rm K}^{2}_{\rm E} \big\}, \left\{ {\rm LP}, {\rm K}^{1}_{\rm O}, {\rm K}^{2}_{\rm O}, {\rm K}^{3}_{\rm O} \right\}, \; \text{and} \;
\left\{ {\rm LP}, {\rm K}^{1}_{\rm E}, {\rm K}^{2}_{\rm E}, {\rm K}^{3}_{\rm E} \right\}. \nonumber
\end{eqnarray} 
If the zero-energy manifold gets split by lifting the SU(2) spin rotational symmetry, it can be accompanied by developing QSHI ($\vec{\rm SH}$) or two spin-triplet Kekul\'e currents ($\vec{\rm K}_{\rm E}$ and $\vec{\rm K}_{\rm O}$), manifesting the color degeneracy of competing orders near the core of a superconducting vortex.

(5) In the vortex core of spin-singlet $s$-Kekul\'e pairing ($M_1={\rm sK}_1$ and $M_2={\rm sK}_2$), one can find layer antiferromagnet ($\vec{{\rm N}}$), QSHI ($\vec{\rm SH}$), and specific components of spin-singlet (${\rm K}_{\rm E}$) and spin-triplet ($\vec{\rm K}_{\rm O}$) Kekul\'e currents as competing orders. The corresponding four-tuplets of mutually anticommuting masses are 
\allowdisplaybreaks[4]
\begin{eqnarray}
&& \left\{ {\rm SH}^1, {\rm SH}^2, {\rm N}^3, {\rm K}^3_{\rm O} \right\}, \left\{ {\rm SH}^2, {\rm SH}^3, {\rm N}^1, {\rm K}^1_{\rm O} \right\}, \big\{ {\rm SH}^1, {\rm SH}^3, \nonumber \\
&& {\rm N}^2, {\rm K}^2_{\rm O} \big\}, \left\{ {\rm N}^1, {\rm N}^2, {\rm N}^3, {\rm K}_{\rm E} \right\}, \; \text{and} \;
\left\{ {\rm K}^1_{\rm O}, {\rm K}^2_{\rm O}, {\rm K}^3_{\rm O}, {\rm K}_{\rm E} \right\}. \nonumber
\end{eqnarray}
On the other hand, the SU(2) spin rotational symmetry of the zero modes can be lifted by layer antiferromagnet ($\vec{\rm N}$), QSHI ($\vec{\rm SH}$), and spin Kekul\'e current ($\vec{\rm K}_{\rm O}$), manifesting the announced color degeneracy among the competing orders. A similar algebra among ten masses in the vortex of $p$-Kekul\'e superconductor can be constructed after taking ${\rm K}_{\rm O} \to {\rm K}_{\rm E}$ and $\vec{\rm K}_{\rm O} \to \vec{\rm K}_{\rm E}$. Therefore, vortex core of all spin-singlet superconductors (the $s$-wave and two Kekul\'e ones) supports topological QSHI.

(6) Finally, we focus on the vortex phase of the spin-triplet $f$-wave pairing. For concreteness, we choose the spin orientation of the superconducting order parameter along the $z$-direction (easy-axis), i.e., $M_1={\rm F}^3_1$ and $M_2={\rm F}^3_2$. Inside the vortex core, one then finds layer polarized state (LP), easy-plane components of layer anti-ferromagnet (${\rm N}^1$, ${\rm N}^2$), and spin-triplet Kekul\'e currents (${\rm K}^1_{\rm E}$, ${\rm K}^2_{\rm E}$, ${\rm K}^1_{\rm O}$, ${\rm K}^2_{\rm O}$), easy-axis QSHI (${\rm SH}^3$), and singlet Kekul\'e currents (${\rm K}_{\rm E}$ and ${\rm K}_{\rm O}$). The five sets of four mutually anticommuting masses in this case are 
\allowdisplaybreaks[4] 
\begin{eqnarray}
&& \left\{ {\rm N}^1, {\rm SH}^3, {\rm K}^1_{\rm E}, {\rm K}^1_{\rm O} \right\}, \left\{ {\rm N}^2, {\rm SH}^3, {\rm K}^2_{\rm E}, {\rm K}^2_{\rm O} \right\}, \big\{ {\rm K}_{\rm E}, {\rm K}_{\rm O}, \nonumber \\
&&  {\rm N}^1, {\rm N}^2 \big\}, \left\{ {\rm LP}, {\rm K}_{\rm E}, {\rm K}^1_{\rm O}, {\rm K}^{2}_{\rm O} \right\}, 
\; \text{and} \;
\left\{ {\rm LP}, {\rm K}_{\rm O}, {\rm K}^{1}_{\rm E}, {\rm K}^{2}_{\rm E} \right\}. \nonumber
\end{eqnarray}
Therefore, all four gapped superconductors support QSHI and some translational symmetry breaking masses in the vortex core. It is also interesting to notice that the vortex phase of pair density waves additionally supports the N\'{e}el layer antiferromagnet.

    \subsubsection{Skyrmion}~\label{subsubsection:skyrmion_valleyQBT_examples}	

Now we consider skyrmions of triplet or three mutually anticommuting masses in BLG.  

(1) First, consider the skyrmion of N\'{e}el layer anti-ferromagnet, with $M_j={\rm N}^j$ for $j=1,2,3$. The six mass matrices that anticommute with $H^{\rm Nam}_{\rm skyr}$ [see Eq.~(\ref{eq:Nambu_skyrmion})] in this case are Kekul\'e currents (${\rm K}_{\rm E}$ and ${\rm K}_{\rm O}$) and spin singlet Kekul\'e superconductors (${\rm sK}_1$, ${\rm sK}_2$, ${\rm pK}_1$, and ${\rm pK}_2$). The generalized U(1) charge of the skyrmion 
\begin{eqnarray}
Q^{\mbox{N\'eel}}_{\rm skyr}=\tau_3 \otimes \sigma_0 \otimes \eta_3 \otimes \alpha_0,
\end{eqnarray}  
is the chiral or valley charge ($Q_{ch}$), which changes sign between two valley. The valley charge rotates between the following three pairs of masses, (1) ${\rm K}_{\rm E}$ and ${\rm K}_{\rm O}$, (2) ${\rm sK}_1$ and ${\rm pK}_2$, and (3) ${\rm sK}_2$ and ${\rm pK}_1$, manifesting the color degeneracy of competing order near the skyrmion core. One generator of the SU(2) isospin is the electric charge $Q_{\rm elec}=\tau_3 \otimes \sigma_0 \otimes \eta_0 \otimes \alpha_0$, which rotates between the real and imaginary components of the $s$-Kekul\'e (${\rm sK}_1$ and ${\rm sK}_2$) and $p$-Kekul\'e (${\rm pK}_1$ and ${\rm pK}_2$) pair density waves. Therefore, skyrmion core of N\'eel order can become charged by nucleating a specific Kekul\'e superconductor.~\footnote{In Dirac systems, such as MLG, a skyrmion of the N\'eel order does not permit any superconducting mass in its core~\cite{hou-mudry-chamon-ryu}.}

(2) Next, we consider a skyrmion of QSHI, with $M_j={\rm SH}^j$ for $j=1,2,3$. Six competing masses in this case are the real and imaginary components of $s$-wave (${\rm S}_1$, ${\rm S}_2$), $s$-Kekul\'e (${\rm sK}_1$, ${\rm sK}_2$), and $p$-Kekul\'e (${\rm pK}_1$, ${\rm pK}_2$) pairings. The U(1) charge of this skyrmion  
\begin{eqnarray}
Q^{\rm QSHI}_{\rm skyr}=\tau_3 \otimes \sigma_0 \otimes \eta_0 \otimes \alpha_0, 
\end{eqnarray}  
is the regular electric charge ($Q_{\rm elec}$), which rotates between the real and imaginary components of all three spin-singlet pairing masses ($s$ wave and two Kekul\'e pairings). One generator of SU(2) isospin is the chiral charge $Q_{ch}$. Hence, the core of a skyrmion of QSHI can host three different types of spin-singlet superconductors, leading to the notion of the color degeneracy among competing orders. In contrast, only the $s$-wave pairing can be realized in the skyrmion core of QSHI in MLG~\cite{grover-senthil}.


\section{Summary and Discussion}~\label{sec:summary}

To summarize, here we unveil the internal algebra of competing orders inside the core of the topological defects, such as vortices and skyrmions, of various mass ordered phases in two-dimensional fermionic systems that in the normal phase are described by biquadratic touching of the valence and conduction bands. We consider two realizations of such systems, describing singled-flavored and valley-degenerate QBTs, respectively, realized on the checkerboard or Kagome lattice~\cite{sun-fradkin-kivelson} and Bernal-stacked BLG~\cite{graphene-RMP}. In the former system, four zero-energy modes bound to the vortex can be split by three competing masses, while the core of a skyrmion possesses a unique charge and supports a doublet of competing masses. For example, zero modes bound to the vortex of the $s$-wave superconductor get split by the QSHI, while a skyrmion of QSHI becomes electrically charged and sustains $s$-wave pairing in its core.

The internal algebra of the competing orders in valley-degenerate QBT systems (such as BLG) is much richer. For example, eight zero-energy vortex modes can supports ten masses that close an SO(5) algebra. While there are five possible patterns for splitting the zero modes by lifting its SO(4) chiral symmetry, they can also be split by spontaneously breaking the SU(2) chiral symmetry in ten different ways. Most interestingly, each SU(2) symmetry can be broken by three distinct sets of chiral triplet masses, giving rise to the notion of color degeneracy of competing orders inside the vortex core. As a concrete example of such color degeneracy, we note that zero modes bound to the vortex of Kekul\'e current orders can be split by spontaneously breaking the SU(2) spin rotational symmetry by either N\'eel layer antiferromagnet or the real and imaginary components of the spin-triplet $f$-wave pairing.

On the other hand, a skyrmion composed of three mutually anticommuting masses possesses an SU(2)$\otimes$U(1) chiral symmetry, and therefore supports a generalized U(1) charge and SU(2) isospin. While the U(1) charge rotates between three distinct pairs of masses, each generator of SU(2) isospin symmetry rotates between two distinct pairs of masses, once again yielding the color degeneracy among competing orders within the skyrmion core. As a concrete outcome of such a rich algebraic structure, we note that skyrmions of QSHI and N\'eel antiferromagnet supports singlet Kekul\'e pairings, while the vortex phase of spin-singlet pair-density-waves (s-Kekul\'e and p-Kekul\'e) supports both the insulating masses.

A question of practical importance arise quite naturally. How to stabilize a real space vortex in an ordered phase? Notice that an easy-plane configuration of N\'{e}el layer antiferromagnet or topological QSHI can be realized in the presence of an in-plane external magnetic field~\cite{roy-BLG-QHE}, which only couples to the spin of electrons (Zeeman coupling) without causing the Landau quantization~\cite{lu-seradjeh}, restricting these two order parameters within the easy-plane, thereby providing the requisite U(1) symmetry to support a vortex. Superconducting vortex can be realized in BLG by bringing a type-II superconductor, such as Nb, to close proximity and applying a magnetic field ($H$) such that $H_{c1}< H \ll H_{c2}$, where $H_{c1}$ ($H_{c2}$) is the lower (upper) critical magnetic field. On the other hand, a two-component mass order, such as the spin-singlet Kekul\'e current, is expected to support a vortex defect deep inside the ordered phase, where the amplitude gap gets frozen, and gapless excitations can only be developed by the phase of the corresponding two-component order parameter in the form of a vortex.

Deep inside a triplet ordered phase, such as layer antiferromagnet and QSHI, once again the amplitude of the order parameter gets frozen, and its phase degrees of freedom assuming the profile of skyrmions are expected to appear naturally. Recently it has been shown that singlet $s$-wave pairing can be nucleated through the condensation of skyrmions of QSHI in MLG~\cite{assaad-QSHI-SC-Natcomm}.
This mechanism involves multiple skyrmion defects and tunneling between them, ultimately causing a global phase coherence of the induced order, otherwise found locally near each defect core, leading to its uniform condensation. However, the nature of the induced competing orders does not depend on whether we take into account an isolated defect or an ensemble of such topological defects, rather only rely on its internal algebra with the parent state hosting such defects. Furthermore, continuous quantum phase transitions between two distinct broken symmetry phases in the presence of topological WZW terms can now be demonstrated in quantum Monte Carlo simulations within the half-filled zeroth Landau level of MLG, without encountering the infamous sign problem~\cite{assaad-mong-1,assaad-mong-2}. These recent developments are encouraging, and should be applicable for Bernal BLG, where continuous phase transitions driven by charge-WZW and isospin-WZW terms can be tested numerically. Finally, we note that each ordered phase is characterized by distinct experimentally measurable signature, which has been discussed for Bernal bilayer graphene in details in Ref.~\cite{roy-classification}. In the presence of local formation of these orders near the core of topological defects they can be identified from various local probes, such as scanning tunneling microscope.

\acknowledgements

This work was supported by the Startup grant of B.R.\ from Lehigh University and NSF CAREER Grant No.\ DMR-2238679 of B.R. Author thanks Igor F. Herbut for useful discussions and correspondences, and Max Planck Institute for the Physics of Complex Systems, Dresden, Germany for hospitality. The author is thankful to Christopher A.\ Leong and Vladimir Juri\v ci\' c for critical reading of the manuscript.

\appendix

\section{No doubling for quadratic band touching}~\label{append:quadratic-minimal}

Low energy excitations around a QBT point in a 2D Brillouin zone is described by the effective Hamiltonian 
\begin{equation}
H_{\rm QBT} ({\mathbf k})\; = \alpha_1 \; d_1 ({\mathbf k}) + \alpha_2 \; d_2 ({\mathbf k}),
\end{equation}
where $d_j({\mathbf k})$ are defined in Eq.~(\ref{formfactor}), and $\alpha_1$ and $\alpha_2$ are mutually anti-commuting Hermitian matrices with the property $\alpha^2_1=\alpha^2_2=I_n$. But their dimensionality ($n$) remains unspecified for now. If there exists another Hermitian matrix, say $\beta$, which anti-commutes with both $\alpha_1$ and $\alpha_2$, spectral symmetry of the energy eigenvalues is guaranteed. Next we ask the following question. What is the minimum dimensionality of $\alpha_i$s, so that $H_{\rm QBT}({\mathbf k})$ is time-reversal invariant?

Let us assume $\alpha_i$s are two-dimensional matrices. The maximal number of mutually anti-commuting two-dimensional Hermitian matrices is three, and they close a $C(3,0)$ algebra. Two of them are purely real, while the remaining one is purely imaginary. One can immediately identify them as the Pauli matrices. Without any loss of generality, we can choose $\alpha_1$ and $\alpha_2$ to be purely real.

 The time reversal symmetry is represented by an anti-unitary operator $I_t = A {\mathcal K}$, where $A$ is a unitary matrix. As we focus on the time-reversal symmetric system, 
\begin{equation}
I_t \: H_{\rm QBT} ({\mathbf k}) \: I^{-1}_t \: =\: H^\star_{\rm QBT} (-{\mathbf k}),
\label{TRSinvarint}
\end{equation}       
since it describes the motion of spinless free fermions on real space. Moreover, for spinless fermions one must have $I^2_t = +1$~\cite{gotfried}. Note that $d_1({\mathbf k})$ and $d_2({\mathbf k})$ do not change sign under the reversal of time. Since we have taken $\alpha_1$ and $\alpha_2$ to be real, Eq.~(\ref{TRSinvarint}) is satisfied when 
\begin{equation}
\left[ A, \alpha_1\right] \:=\: \left[ A, \alpha_2 \right] \:=\: 0. 
\end{equation} 
For two-dimensional matrices, there exist only one matrix which commutes with all the three mutually anticommuting Pauli matrices, the identity matrix ($\sigma_0$) with its trace being equal to 2. The time-reversal operator is, therefore, $I_t = {\mathcal K}$, and $I^2_t = + 1$, which is independent of the choice of basis. Therefore, when valence and conduction band display a quadratic touching, the minimal representation of such a system can be two-component, and therefore the system does not necessarily encounter the \emph{fermion doubling}. On the other hand, when $d_1({\bf k})= v k_x$, and $d_2({\bf k})=v k_y$, where $v$ is the Fermi velocity, the minimal representation of $\alpha_1$ and $\alpha_2$ is four-dimensional for spinless fermions in time-reversal symmetric Dirac systems~\cite{herbut-minimalrepresentation}, which leads to the notion of fermion doubling for chiral Dirac fermions, such as in MLG, according to the Nielsen-Ninomiya theorem~\cite{nielsen-ninomiya}.


\section{Generators of SO(5) and SO(4)}~\label{append:so5so4}

Ten generators of an SO(5) group can be labeled as $J_{\alpha \beta}$, where $\alpha, \beta=2, \cdots, 6$, which satisfy the anti-symmetric property $J_{\alpha \beta}=-J_{\beta \alpha}$. In addition, they satisfy the following commutation relation 
\begin{align}~\label{eq:so5algebra}
\left[ J_{\alpha \beta}, J_{\mu \nu} \right]= i \big[ \delta_{\beta \mu} J_{\alpha \nu} + \delta_{\alpha \nu} J_{\beta \mu} 
- \delta_{\beta \nu} J_{\alpha \mu} - \delta_{\alpha \mu} J_{\beta \nu} \big].
\end{align}
In order to show that ten generators from Eq.~(\ref{eq:so5generator}), close an SO(5) algebra we write the first four entries of ${\mathcal G}$ as 
\begin{equation}
E M_j= \left( \Gamma_1 \Gamma_2 M_1 \right) M_2 M_j \equiv 2 J_{2j},
\end{equation}   
such that $J_{2j}=-J_{j2}$ for $j=3,4,5,6$. The rest of the six entries from ${\mathcal G}$ can be expressed as $E_{jk}=2J_{jk}$ for $j=3,4,5,6$, and they also satisfy the antisymmetry property. Now it is straightforward to show that ten generators appearing in ${\mathcal G}$, expressed as $J_{\alpha \beta}$, where $\alpha, \beta=2, \cdots, 6$, satisfy the commutation relation in Eq.~(\ref{eq:so5algebra}).

Next we show that five sets of six generators appearing in Fig.~\ref{so4_squares}(a)-\ref{so4_squares}(e) (in blue) close SO(4) algebra. For concreteness, we focus on the six generators appearing in Fig.~\ref{so4_squares}(a). Following the same steps, one can show that other four sets of six generators also close SO(4) algebra. An SO(4) group has six generators, namely 
\begin{equation}
\boldsymbol{A}=\left( A_1, A_2, A_3 \right),
\: \text{and} \:
\boldsymbol{B}=\left( B_1, B_2, B_3 \right),  
\end{equation}  
satisfying the commutation relations
\begin{equation}~\label{eq:so4commutation}
\left[ A_j, A_k \right]=i \epsilon_{jkl} A_l, 
\left[ B_j, B_k \right]=i \epsilon_{jkl} B_l, 
\left[ A_j, B_k \right]=i \epsilon_{jkl} B_l,
\end{equation}
for $j,k,l=1,2,3$. From the six generators appearing in Fig.~\ref{so4_squares}(a), we choose
\begin{eqnarray}
A_1 &=& -\frac{E_{34}}{2}, A_2=-\frac{E_{45}}{2}, A_3=-\frac{E_{53}}{2}, \nonumber \\
B_1 &=&  \frac{E_{56}}{2}, B_2= \frac{E_{36}}{2}, B_3= \frac{E_{46}}{2}.
\end{eqnarray} 
It is now straightforward to show that for these choices of $\boldsymbol{A}$ and $\boldsymbol{B}$, the commutation relations from Eq.~(\ref{eq:so4commutation}) are satisfied. To show that the Lie algebra of the SO(4) group is isomorphic to that of SU(2)$\otimes$SU(2) group, we construct six new generators according to 
\begin{equation}
X_j=\frac{1}{2} \left( A_j + B_j \right), \; Y_j=\frac{1}{2} \left( A_j - B_j \right),
\end{equation}
for $j=1,2,3$. It is now straightforward to show that individually $\boldsymbol{X}$ and $\boldsymbol{Y}$ close SU(2) algebra, but these two sets of three generators commute with each other, i.e., 
\begin{equation}
\left[ X_j, X_k \right]=i \epsilon_{jkl} X_l,
\left[ Y_j, Y_k \right]=i \epsilon_{jkl} Y_l,
\left[ X_j, Y_k \right]=0,
\end{equation}
for $j,k,l=1,2,3$. Therefore, $\boldsymbol{X}$ and $\boldsymbol{Y}$ are the generators of two decoupled SU(2), and at the level of Lie algebras SO(4)$\cong$SU(2)$\otimes$SU(2).


\end{document}